\documentclass[superscriptaddress, aps, pra, 10pt, nofootinbibace]{revtex4-2}

\usepackage{tabularx}
\usepackage{xcolor}
\usepackage{bm}
\usepackage{comment}
\usepackage{graphicx}
\usepackage{multirow}
\usepackage{array}
\usepackage{booktabs} 
\usepackage{tabularray}
\usepackage[normalem]{ulem}
\usepackage{amsmath}
\usepackage{mathtools}
\usepackage{amssymb}
\usepackage{acro}
\usepackage[utf8]{inputenc}
\usepackage[T1]{fontenc}
\usepackage{url}
\usepackage{verbatim}
\usepackage{mdframed}
\usepackage{physics}
\usepackage{mathrsfs}
\usepackage{subcaption}

\DeclareUnicodeCharacter{0393}{$\Gamma$}

\DeclareAcronym{DFT}{
short = DFT,
long = Density-Functional Theory
}

\DeclareAcronym{SOAP}{
short = SOAP,
long = Smooth Overlap of Atomic Positions
}

\DeclareAcronym{ACE}{
short = ACE,
long = Atomic Cluster Expansion
}

\DeclareAcronym{MACE}{
short = MACE,
long = Multi-ACE
}

\DeclareAcronym{RR}{
short = RR,
long =  Ridge Regression
}

\DeclareAcronym{KRR}{
short = KRR,
long = Kernel Ridge Regression
}

\DeclareAcronym{GAP}{
short = GAP,
long = Gaussian Approximation Potential
}

\definecolor{yscol}{HTML}{6622AA}

\def\R{\mathbb{R}}

\def\RR{\bm{R}}

\def\ACEstar{ACE~($\ast$)}

\newcommand{\thetasc}{\theta_\text{sc}}

\begin{document}

\title{An Atomic Cluster Expansion Potential for Twisted Multilayer Graphene}

\author{Yangshuai Wang}
\email{yswang@nus.edu.sg}
\affiliation{Department of Mathematics, National University of Singapore, 10 Lower Kent Ridge Road, Singapore 119076.}

\author{Drake Clark}
\email{clar1939@umn.edu}
\affiliation{School of Mathematics, University of Minnesota, Minnesota, USA 55455.}

\author{Sambit Das}
\email{dsambit@umich.edu}
\affiliation{Department of Mechanical Engineering, University of Michigan, Ann Arbor, Michigan, USA 48109}

\author{Ziyan Zhu}
\email{ziyan.zhu@bc.edu}
\affiliation{Stanford Institute for Materials and Energy Sciences,
SLAC National Accelerator Laboratory, 2575 Sand Hill Road, Menlo Park, CA 94025, USA}
\affiliation{Department of Physics, Boston College, Chestnut Hill, MA 02467, USA}

\author{Daniel Massatt}
\email{dmassatt@lsu.edu}
\affiliation{Department of Mathematics, Louisiana State University, USA 70803}

\author{Vikram Gavini}
\email{vikramg@umich.edu}
\affiliation{Department of Mechanical Engineering, Department of Materials Science \& Engineering, University of Michigan, Ann Arbor, Michigan, USA 48109}

\author{Mitchell Luskin}
\email{luskin@umn.edu}
\affiliation{School of Mathematics, University of Minnesota, Minnesota, USA 55455.}

\author{Christoph Ortner}
\email{ortner@math.ubc.ca}
\affiliation{Department of Mathematics, University of British Columbia, 1984 Mathematics Road, Vancouver, BC, Canada V6T 1Z2.}

\date{\today}

\begin{abstract}
{
 Twisted multilayer graphene, characterized by its moir\'e patterns arising from inter-layer rotational misalignment, serves as a rich platform for exploring quantum phenomena. Machine learning interatomic potentials (MLIPs) are a promising approach to model such systems. Our work develops a method to generate training and test datasets for fitting MLIPs that capture all possible misalignments but remain small-scale to facilitate efficient data generation and parameter estimation. To achieve this, we generate configurations with periodic boundary conditions suitable for DFT calculations, and then introduce an internal twist and shift within those supercell structures. Using this technique, supplemented with an active learning workflow, we fit an Atomic Cluster Expansion potential for simulating twisted multilayer graphene and test it for accuracy and robustness on a range of simulation tasks.
}
\end{abstract}

\maketitle

\section{Introduction}
\label{sec:introduction}
Multilayer graphene has garnered significant attention due to its unique electronic properties and potential applications in next-generation devices~\cite{Novoselov2012, Bonaccorso2010}. { When graphene layers are stacked with a slight rotational misalignment, known as a twist angle, they form moir\'e patterns, leading to novel physical phenomena~\cite{Cao2018, Bistritzer2011, marvels, simulator, twistronics}. Understanding these phenomena is crucial for advancing condensed matter physics and developing materials with tunable electronic characteristics.} The ability to manipulate these properties through the twist angle makes the twisted graphene systems a versatile platform for exploring quantum mechanical effects. Consequently, research into twisted bilayer and multilayer graphene holds great promise for innovations in nanoelectronics, photonics, and quantum computing.

First-principles calculations of material properties offer a robust method for discovering new materials and optimizing existing ones. However, these calculations are resource-intensive and their applicability is often limited. This limitation is especially pronounced in multilayer twisted materials, which do not always conform to periodic boundary conditions, rendering first-principles methods such as density functional theory (DFT) computationally infeasible. Empirical interatomic potentials (IPs), including Lennard-Jones~\cite{wang2020lennard}, the embedded atom method (EAM)~\cite{daw1993embedded}, Stillinger-Weber~\cite{vink2001fitting}, and Kolmogorov-Crespi \cite{KolmogorovCrespi2005,Hod2010} provide significant computational speed-ups—typically by six to eight orders of magnitude compared to DFT. These potentials, with their straightforward and physically motivated forms, offer reasonable predictions for low-energy structures but generally fall short in achieving the quantitative accuracy of DFT, particularly when reproducing macroscopic properties.

To overcome these limitations, machine-learning interatomic potentials (MLIPs) have emerged as a transformative approach in molecular and materials simulation~\cite{Behler2007ACSF, bartok2010gaussian, THOMPSON2015SNAP, MTP2016, ACE_ralf}. MLIPs enable the simulation of atomistic systems at or near the accuracy of electronic structure methods while being computationally cheaper by orders of magnitude. This capability has made the simulation of large-scale systems and long time-scales at high accuracy accessible, establishing MLIPs as indispensable tools for atomic-scale simulation. Recent comprehensive reviews of the field are provided in~\cite{2019_deringer_review, behler_csanyi_2021, DeringerCsanyi2021ChemRev, Ceriotti2021Review}. Of particular relevance to the present work are the atomic cluster expansion and related methods introduced in~\cite{bartok2010gaussian, THOMPSON2015SNAP, MTP2016, ACE_ralf}. The creation of an MLIP begins with a flexible functional form, constrained only by physical requirements in which there is high confidence, for example, to comply with  the natural symmetries of the potential energy in three-dimensional space. Parameters are then estimated using reference data, typically energies, forces, and virial stresses from a representative set of atomic configurations. This data is usually generated via quantum mechanical techniques, such as DFT calculations, which are often limited to relatively small structures. A well-trained MLIP is expected to provide accurate predictions for processes on both similar and much larger spatial scales. 

The Atomic Cluster Expansion (ACE), introduced in~\cite{ACE_ralf}, is a particularly flexible and theoretically sound MLIP approach. ACE offers a balance between cost and accuracy and has been successfully applied across a wide range of tasks, particularly in materials simulation~\cite{2019_Seko, DUSSON2022, performant2022lysogorskiy, hyperactive2022, 2023-ace_PtRh, 2022_pacemaker, 2022_drautz_ace_C}. Linear variants of the ACE model have proven to be both data and computationally efficient, making them particularly useful in active learning (AL) workflows~\cite{hyperactive2022}. Linearity also facilitates sensitivity analysis and paves the way for reliable uncertainty quantification. Throughout this work, we adopt the ACE method, but the methods and benchmarks we introduce are likely applicable to the vast majority of MLIPs frameworks.

Although MLIPs have been extensively utilized in various fields, their application to twisted multilayer graphene systems remains largely unexplored. { Previous studies have developed MLIPs for other two-dimensional (2D) materials, such as MoS$_2$, monolayer graphene, buckled silicone, and quaternary transition metal dichalcogenide (TMD) alloy, to investigate thermal conductivity and phononic properties~\cite{mortazavi2020exploring, mortazavi2020efficient, marmolejo2022thermal, kocabacs2023gaussian, anas2024}. For instance, Magorrian {\em et al.}~\cite{magorrian2024strong} focused on atomistic relaxation in twisted bilayers of InSe, while Jung {\em et al.}~\cite{jung2023artificial} explored mechanical and fracture dynamics in monolayer graphene. Yin {\em et al.}~\cite{yin2024machine} applied machine-learning force fields for membrane design, and Wilson {\em  et al.}~\cite{wilson2022batch} used batch active learning to construct MLIPs, emphasizing phonon dispersion and phase transition temperature calculations for monolayer GeSe. More recently, Bao \textit{et al.}~\cite{bao2025transfer} leveraged transfer learning to predict the electronic structure of twisted MoTe$_2$, while a closely related study~\cite{dpmoire2024} developed a framework for constructing MLIPs for moir\'e systems, specifically MX$_2$ (M = Mo, W; X = S, Se, Te) materials. Georgaras \textit{et al.}~\cite{georgaras2025accurate} further advanced this direction by proposing a split MLIP architecture and a curated dataset strategy that explicitly distinguishes intralayer and interlayer interactions, leading to improved accuracy in modeling moir\'e materials.} These studies provide a general, theoretically and computationally attractive approach to predicting physical properties on homogeneous structures, such as relaxed lattice constants, phonon spectra, and band structures. 


In the present study, we develop methods to expand the training set and extensive benchmarking of MLIPs for twisted multilayer systems in- and out-of-distribution; to be used in a stand-alone fashion or integrated into more general frameworks such as DPmoire~\cite{dpmoire2024}.
Specifically, we develop a training and benchmark protocol for twisted multilayer graphene systems and apply it to fitting an ACE model. A key challenge in developing an MLIP is the selection of small-scale training structures. For twisted multilayer graphene this is particularly challenging since most twist angles of interest cannot be represented in a small periodic structure accessible to DFT methods. A novel contribution of our work to overcome this is to employ locally twisted training structures. This provides a systematic method to generate training configurations with periodic boundary conditions that are suitable for DFT calculations across all twist angles. We combine this approach with Bayesian active learning and filtering techniques that sub-select configurations based on uncertainties derived from Bayesian linear regression. 


Our evaluation of the developed models includes several numerical experiments designed to test the effectiveness of our method compared with naively generated datasets, as well as the the in-distribution accuracy and out-of-distribution robustness of the resulting fitted models. Our tests include geometry optimization, focusing on defect formation and migration within the twisted layers,  molecular dynamics (MD) simulations including static, dynamic, and thermodynamic limits. These simulations provided insights into the material's stability, response to thermal fluctuations, and overall dynamical behavior. Additionally, we performed phonon calculations to explore the vibrational properties of twisted multilayer graphene, which is crucial for understanding its thermal conductivity and other phonon-related phenomena. Our tests demonstrate the wide applicability and effectiveness of our approach to training MLIPs for twisted multi-layer structures.

\section{Methods}
\label{sec:methods} 
\subsection{Multilayer twisted graphene}
\label{sec:sub:multilayerTBG}
\begin{figure}
    \centering
    \includegraphics[width=0.32\linewidth]{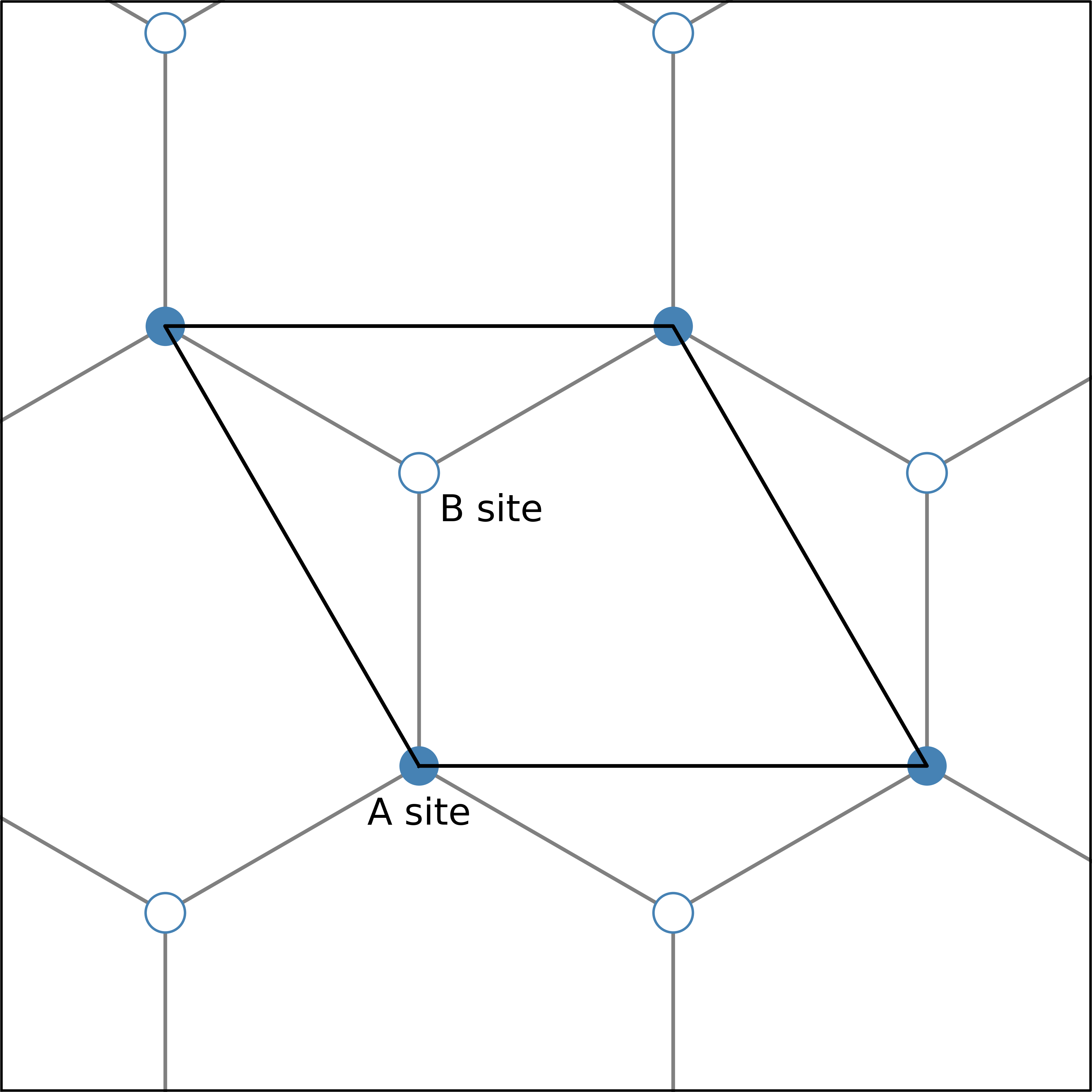}
    \includegraphics[width=0.32\linewidth]{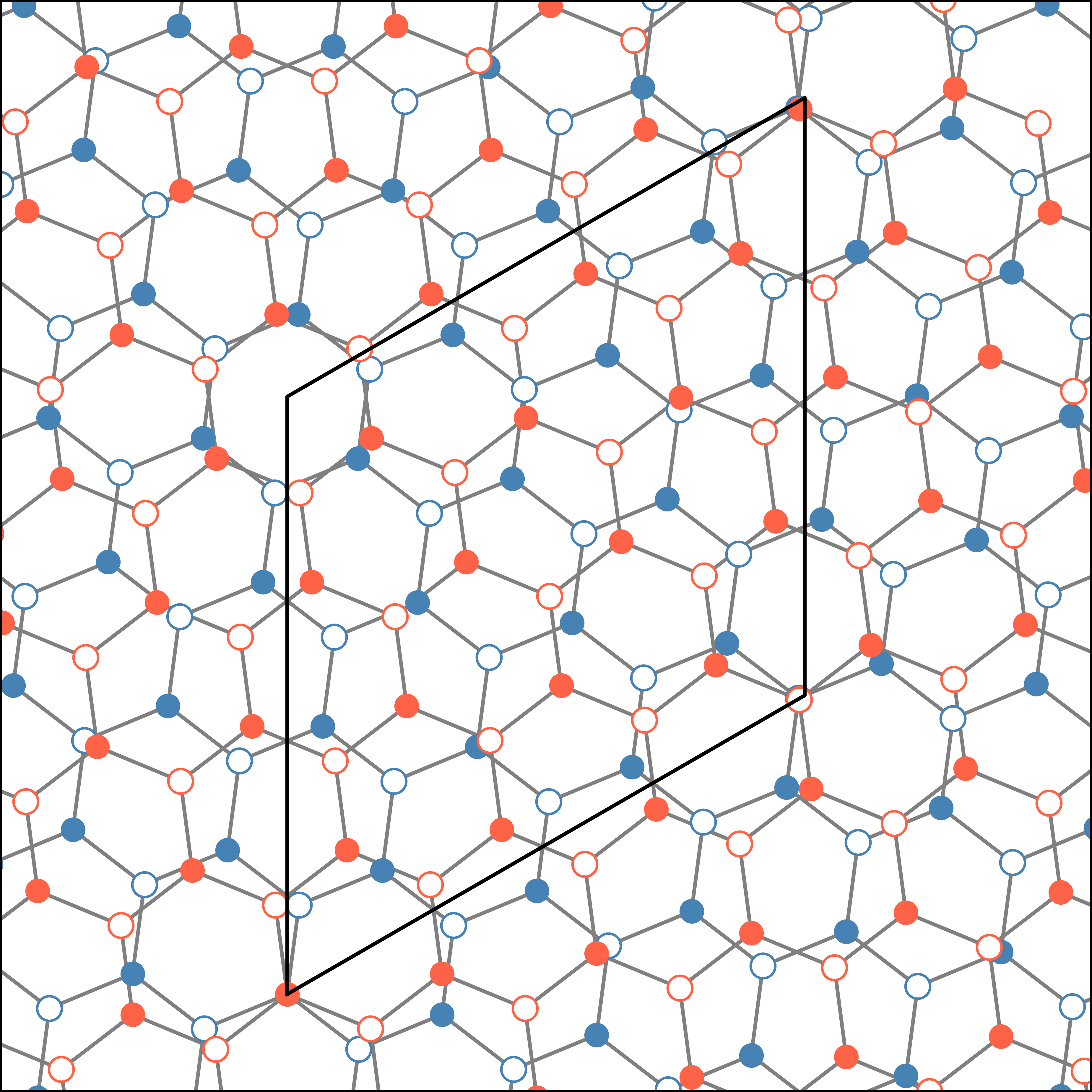}
    \includegraphics[width=0.32\linewidth]{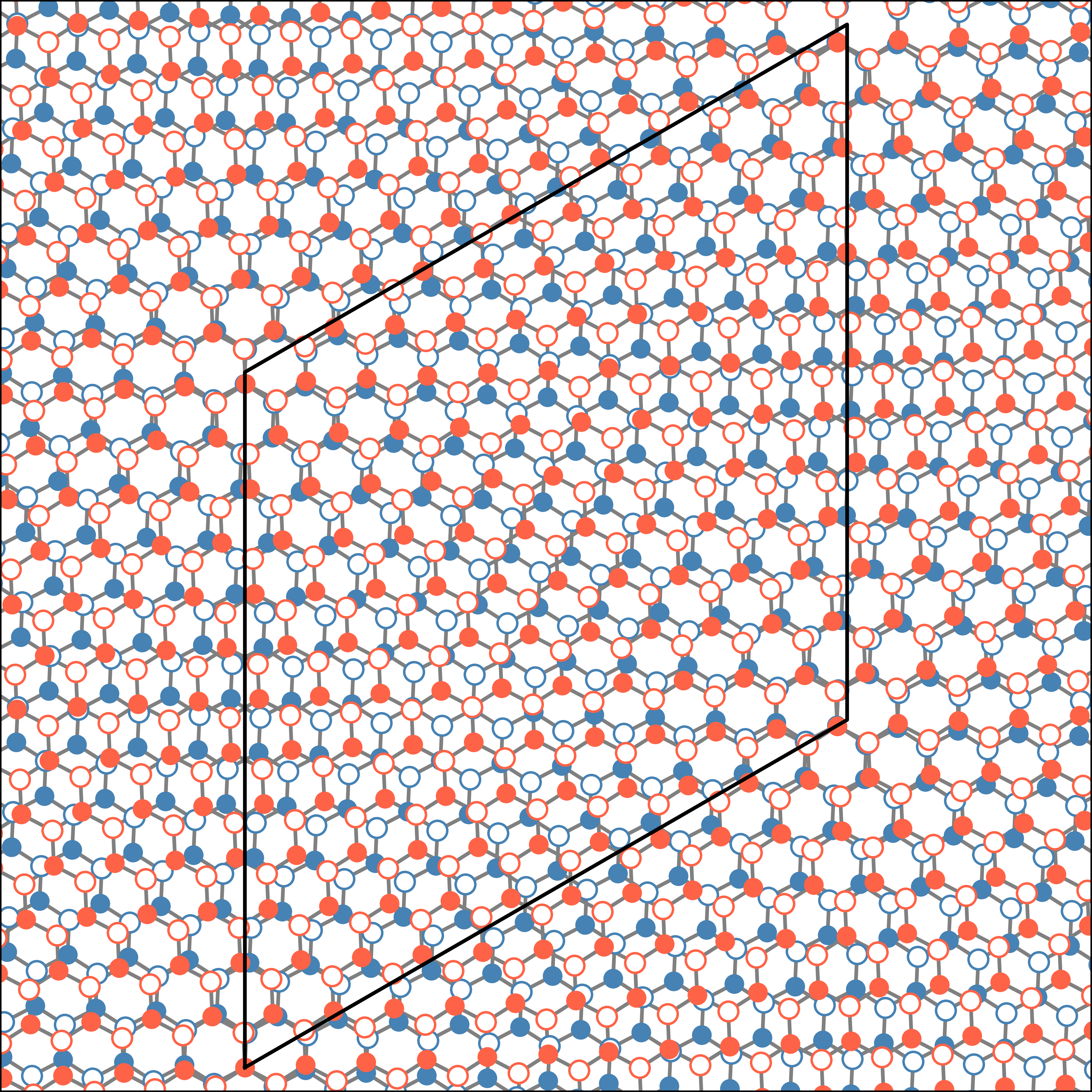}
    \caption{Graphene unit cell (left); moir\'e cell for 15$^\circ$ twisted bilayer graphene (center); moir\'e cell for 5$^\circ$ twisted bilayer graphene (right).}
    \label{fig:graphene_unit_cell}
\end{figure}

%
%
We focus on multilayer twisted graphene, but the methodologies we develop for this case are readily extendable to other layered 2D materials.  A single layer of graphene has a honeycomb structure (see Figure \ref{fig:graphene_unit_cell} left) associated with a lattice whose fundamental matrix is
\begin{equation}\label{eq:gfundmat}
	A=a_0\begin{pmatrix}1 & -1/2 \\ 0 & \sqrt{3}/2 \end{pmatrix} .
\end{equation}
Each unit cell of graphene has two carbon atoms with shifts $b_A=(0,0)^T$ and $b_B=a_0(0,\sqrt3/3)^T$, labeled ``A site'' and ``B site'' in Figure \ref{fig:graphene_unit_cell}.  For our numerical results we used the lattice constant $a_0=2.47$ \AA{, which is taken from experimental measurements as reported in~\cite{2DPerturb15}.} 
Twisted bilayer graphene is formed by stacking two layers of graphene on top of each other with a relative twist.  The two layers' fundamental matrices are then:
\begin{equation}\label{eq:tbgfundmat}
	A_1=R_{-\theta/2}A,\quad A_2=R_{\theta/2}A, \quad \text{with } R_\phi=\begin{pmatrix}\cos\phi & -\sin\phi \\ \sin\phi & \cos\phi \end{pmatrix}.
\end{equation}
The misalignment between the two layers caused by the twist angle produces a large-scale moir\'e pattern. 
The resulting bilayer structure is generally incommensurate and only exhibits approximate periodicity. Exact periodicity is achieved only at specific twist angles where $\cos \theta$ satisfies a Diophantine condition \cite{2DPerturb15,Carr2020}.
The moir\'e unit cell has the form \cite{Cazeaux2019,relaxphysics18}:
\begin{equation}\label{tbgmoire}
  A_M:=(A_1^{-1}-A_2^{-1})^{-1}, \quad A_M(\theta)=\frac{a_0}{2\sin(\theta/2)}\begin{pmatrix}0 & \frac{\sqrt{3}}{2}\\-1 & \frac{1}{2}\end{pmatrix}.
\end{equation}
The moir\'e superlattice length depends on the rotation angle between two layers $\theta$, growing as $\sim 1/\theta$ as $\theta$ decreases.  
Techniques that rely on periodic boundary conditions or supercell approximations require impractically large domains for numerical purposes. This effect is even more severe in the multi-layer case~\cite{zhu2019moir}. 

The large length scale of moir\'e patterns and lack of periodicity make numerical simulation of many interesting phenomena in the small-angle regime challenging. 
A wide variety of strongly correlated electronic phases of matter such as superconductivity~\cite{cao2018unconventional} and fractional quantum anomalous Hall (fQAHE) states~\cite{park2024, cai2024} can emerge. 
Mechanical properties, which are the primary focus of the present paper, including structural relaxation~\cite{yoo2019,relaxfosdick22,KimRelax18,dai2016twisted,relaxphysics18,koshinorelax17}, phonons~\cite{koshinophonon19,zhuphonon22}, and defects, play a critical role in understanding the electronic behavior of these moir\'e systems.
For example, mechanical relaxation can form large triangular or hexagonal domains~\cite{yoo2019}, modifying electronic structures and shifting the twist angle at which fQAHE is expected to appear. Additionally, strong electron-phonon coupling, which has been measured~\cite{birkbeck2024measuring}, may contribute to the observed moir\'e superconductivity.
Defects are also prevalent in these moir\'e systems and their impact on electronic properties is not yet understood.

Mechanical relaxation of multilayer graphene systems has in the past been performed using linear elasticity approximations or the REBO interatomic potential for the intralayer component and the generalized stacking fault energy or the Kolmogorov-Crespi interatomic potential \cite{KolmogorovCrespi2005} for the interlayer component \cite{Wijk2015, relaxphysics18, zhu2019moir,  dai2016twisted}.  Although the Kolmogorov-Crespi interatomic potential \cite{KolmogorovCrespi2005} is a major improvement over two-body interatomic potentials such as Lennard-Jones, there are significant errors in the prediction of forces of some important local configurations (disregistries) when compared to DFT computations ~\cite{tadmorpotential18}. An important perspective of our present work is to develop a methodology (in the simplified multi-layer graphene setting) that can be readily extended to systematically construct accurate interatomic potentials for more complex 2D materials such as the transition metal dichalcogenides (TMDs). 

\subsection{The linear ACE framework}
\label{sec:sub:ACE}
Accurate energy and force calculations are best achieved through electronic structure techniques such as DFT. However, DFT has limitations due to computational cost for large systems and long timescales. To overcome these limitations, MLIPs, trained on DFT data, have become essential in computational materials science due to their high accuracy and transferability, but relatively low computational cost~\cite{2019_deringer_review, behler_csanyi_2021, DeringerCsanyi2021ChemRev, Ceriotti2021Review}. 

\def\RR{\mathbf{A}}

\def\Es{\varepsilon}

We utilize the linear atomic cluster expansion (ACE) parametrization~\cite{witt2023otentials, DUSSON2022, ACE_ralf}. MLIPs for materials typically express the total energy as a sum over individual site energies. To that end, we consider $N$ atoms described by their position vectors $\boldsymbol{r}_j$. A set $\RR := \{\boldsymbol{r}_1, \ldots, \boldsymbol{r}_N\} \in \mathbb{R}^{3N}$ of $N$ particle positions is called an atomic configuration. We suppress the specification of a unit cell and periodicity for the sake of simplicity of presentation. One normally also accounts for the chemical elements, but since this work is only concerned with models for multi-layer graphene, we can ignore this dependency as well.

Let $\boldsymbol{r}_{ij} = \boldsymbol{r}_j - \boldsymbol{r}_i \in \mathbb{R}^3$ be the relative position vector between atom $j$ and a reference atom $i$ and $r_{ij} = |\boldsymbol{r}_{ij}|$ the distance. 
Given a cutoff radius $r_{\text{cut}}$ (the first model hyperparameter), we define the {\it local} atomic environment as $\RR_i := \{\boldsymbol{r}_{ij}\}_{r_{ij} \leq r_{\rm cut}}$.

The total energy of a structure of this kind is decomposed into dispersion interaction and short-ranged many-body site energies
\begin{equation}\label{eq:spatial_decomp}
   E(\RR; \mathbf{c}) = E_{\rm disp}(\RR) + \sum_{i=1}^{N} \Es(\RR_i; \mathbf{c}),
\end{equation}
where the dispersion interaction $E_{\rm disp}$ will always be the D3 dispersion model~\cite{DFTD3} and the site energy $\Es$ is a ``learnable'' function of an atomic environment $\RR_i$. It's parameters  $\mathbf{c}$ will be determined through regression. While the vast majority of local MLIPs are written in the form \eqref{eq:spatial_decomp}, the ACE model we employ has the additional feature that each site energy is explicitly expanded into many-body terms, 
\begin{equation}\label{eq:many-body}
\Es(\RR_i; \mathbf{c}) = 
    \sum_{\nu = 0}^{\nu_{\rm max}} \frac{1}{\nu!} 
    \sum_{j_1, \dots, j_\nu} 
    V_\nu(\boldsymbol{r}_{i j_1}, \dots, \boldsymbol{r}_{i j_\nu}; \mathbf{c}).
\end{equation}
The repeated indices in the summation are deliberate and result in a highly computationally efficient implementation (cf. Appendix~\ref{sec:apd:ACE}).  After further expanding each potential $V_\nu$ in a basis, the ACE site energy $\Es$ can then finally be expressed as 
\begin{equation}\label{eq:params-E}
\Es(\RR_i; \mathbf{c}) := \mathbf{c} \cdot \mathcal{B}(\RR_i) =  \sum_{B \in \mathcal{B}} c_B {B}(\RR_i), 
\end{equation}
with basis functions $B$ and parameters $\mathbf{c}:=\{c_B\}_{B \in \mathcal{B}}$ that are optimized via linear least squares (cf. \eqref{eq:quad-loss}).

Both total energy $E$ and site energy $\Es$ must be invariant under permutations, rotations, and reflections of the inputs; hence the basis $\mathcal{B}$ is constructed to satisfy the same invariance exactly. In addition, the basis functions inherit the natural ordering of \eqref{eq:many-body}, providing a physically interpretable hyper-parameter $\nu_{\rm max}$ to converge the fit accuracy. A more detailed review of the ACE model and its (hyper-)parameters is given in Appendix~\ref{sec:apd:ACE}.

To estimate the model parameters ${\bf c}$ we require a training dataset which contains a list of atomic configurations $\mathfrak{A}=\{\mathbf{A}\}$ and corresponding observations evaluated using reference electronic structure models (we employ DFT; cf. Appendix~\ref{sec:apd:supp} for details). For each configuration $\mathbf{A}$, the total energy $E^{\rm ref}(\mathbf{A}) \in \R$, forces (negative energy gradient) $F^{\rm ref}:=-\nabla E^{\rm ref}(\mathbf{A}) \in \R^{N \times 3}$ (where $N$ is the total number of atoms in each configuration $\mathbf{A}$) and virial stress $\Sigma^{\rm ref}(\mathbf{A})\in\R^{3 \times 3}$ are included. We use a regularized linear least squares method to estimate the parameters, requiring minimization of the quadratic loss function:
\begin{align} \label{eq:quad-loss}
    \mathcal{L}(\mathbf{c}) := \sum_{\mathbf{A} \in \mathfrak{A}} \Big( N_\RR^{-1} w^{2}_{E} \big| E(\mathbf{A}; \mathbf{c}) - E^{\rm ref}(\mathbf{A}) \big|^{2} + w^{2}_{F} \big| F(\mathbf{A}; \mathbf{c}) - F^{\rm ref}(\mathbf{A}) \big|^{2} & \nonumber \\
    + N_\RR^{-1} w^{2}_{\Sigma} \big| \Sigma(\mathbf{A}; \mathbf{c}) - \Sigma^{\rm ref}(\mathbf{A}) \big|^{2} \Big)&\, + \lambda \| \Gamma {\bf c} \|^2,
\end{align}
where $N_\RR$ is the number of atoms in structure $\RR$, $F(\RR; {\bf c})$ is a vector of forces, $\Sigma(\mathbf{A}; \mathbf{c})$ is the virial stress of the ACE model, $\Gamma$ specifies the form of the regularizer (smoothness prior), and $\lambda$ is a scaling parameter that determines the relative weight of regularization. The weights $w_{E}$, $w_{F}$ and $w_{\Sigma}$ adjust the contributions of energy, forces and virial stresses, respectively. Within the Bayesian interpretation of ridge regression (also used as the solver, see the discussion below), $\Gamma$ defines a prior distribution for the model parameters $\mathbf{c} \sim \mathcal{N}({\bf 0}, \Gamma^{-2})$. For details, see Appendix~\ref{sec:apd:blr} and~\cite{witt2023otentials}. Throughout the present work, we choose an algebraic smoothness prior of strength $p=5$, as defined in~\cite[Eq.10]{witt2023otentials}. 

Due to the linearity of the ACE model~\eqref{eq:params-E}, minimizing the quadratic loss~\eqref{eq:quad-loss} is equivalent to solving a linear least squares problem. Various solvers are available for this task; see~\cite{witt2023otentials} for an overview and benchmark tests. 
In this work, we employ Bayesian linear regression to solve the problem, as it automatically selects the regularization parameter $\lambda$ and produces a {\it posterior} parameter distribution that we can then employ for uncertainty quantification. 
We briefly review the approach with detailed discussions deferred to Appendix~\ref{sec:apd:blr}. 

By assuming an isotropic Gaussian prior on the model parameters and Gaussian independent and identically distributed (i.i.d) noise on observations, we obtain an explicit posterior distribution $\text{post}(\mathbf{c})$ for the parameters, from which the standard deviation of the posterior-predictive distribution of total energies (and other quantities) can be deduced.
%
%
Using the posterior directly is computationally expensive for a large basis; scaling as $O(N_{\text{basis}}^2)$, hence we generate an ensemble of parameters $\{\mathbf{c}^k\}^K_{k=1}$ obtained by sampling from the posterior ${\rm post}(\mathbf{c})$, where $K$ is the number of sample (or, committee size). Each parameter vector $\mathbf{c}^k$ results in predicted energies $E^k$ (as well as forces, virials, etc).
%
(See Appendix~\ref{sec:apd:blr} and \cite{hyperactive2022, witt2023otentials} for further details.) This results in an energy uncertainty estimate
\begin{equation}
\label{eq:com_E}
\sigma^{\rm E} := \sqrt{\frac{1}{\beta w^{2}_{E}} + \frac{1}{K} \sum_{k=1}^K (E^k - \overline{E})^2},
\end{equation}
where $\beta$ is a hyperparameter (see Appendix~\ref{sec:apd:blr} for details), $\overline{E}:=\sum_{i=1}^{N}\sum_{B \in \mathcal{B}} \bar{c}_B \cdot B(\RR_i)$
with $\bar{\mathbf{c}}$ 
being the posterior mean of the distribution and $E^k$ denoting the total ACE energy evaluated using $\mathbf{c}^k$.  This approach is computationally efficient, requiring only a single basis evaluation followed by $K$ dot products with the ensemble parameters. Similarly, the corresponding relative force uncertainty for the $i$-th atom, as defined in \cite{hyperactive2022}, is given by:
\begin{equation}
\label{eq:com_F}
\sigma^{\rm F}_i := \frac{1}{K} \sum_{k=1}^{K} \frac{\| F_i^k - \overline{F}_i \|}{\|\overline{{F}}_i\| + \varepsilon},
\end{equation}
where $\overline{{F}}_i$ is the mean force prediction for the $i$-th atom, and $\varepsilon$ is a regularization constant to prevent divergence, set to 0.2 eV/\AA. In the following section, we use $\max_i \sigma^{\rm F}_i$, the maximum force uncertainty among all atoms in a given configuration, as the criterion for selecting a candidate dataset that is as small as possible while still accurately capturing the properties of interest.

\subsection{Dataset generation}
\label{sec:sub:data}

In this section, we describe the process of generating the training and test datasets used to fit and validate the potentials employed in benchmarking twisted multilayer graphene.

\subsubsection{Locally twisted multi-layers}
\label{sec:sub:sub:local}
As mentioned in \S~\ref{sec:sub:multilayerTBG}, bilayer graphene systems with twist angle $\theta$ are only periodic for specific values of $\theta$.  We used periodic supercells with twist angles $\theta=\phi_2-\phi_1$ for $\phi_1$ and $\phi_2$ satisfying 
%
\begin{equation}
	\cos(\phi_1)=\frac{p+2q}{2n},\quad \cos(\phi_2)=\frac{2p+q}{2n},
\end{equation}
where $p,q,n$ are integers satisfying the Diophantine equation \cite{2DPerturb15}
\begin{equation}
	p^2+q^2+p\cdot n=n^2.
\end{equation}
The seven smallest such commensurate systems and their supercell atom count are given in \cite[Table II]{2DPerturb15}.  These supercells quickly become intractable for DFT computations as the angle gets smaller.  There is a more general Diophantine equation that can be solved to form commensurate cells~\cite{Carr2020}, but these still suffer from the same size issue.

\begin{figure}
    \centering
    \includegraphics[width=.3\linewidth]{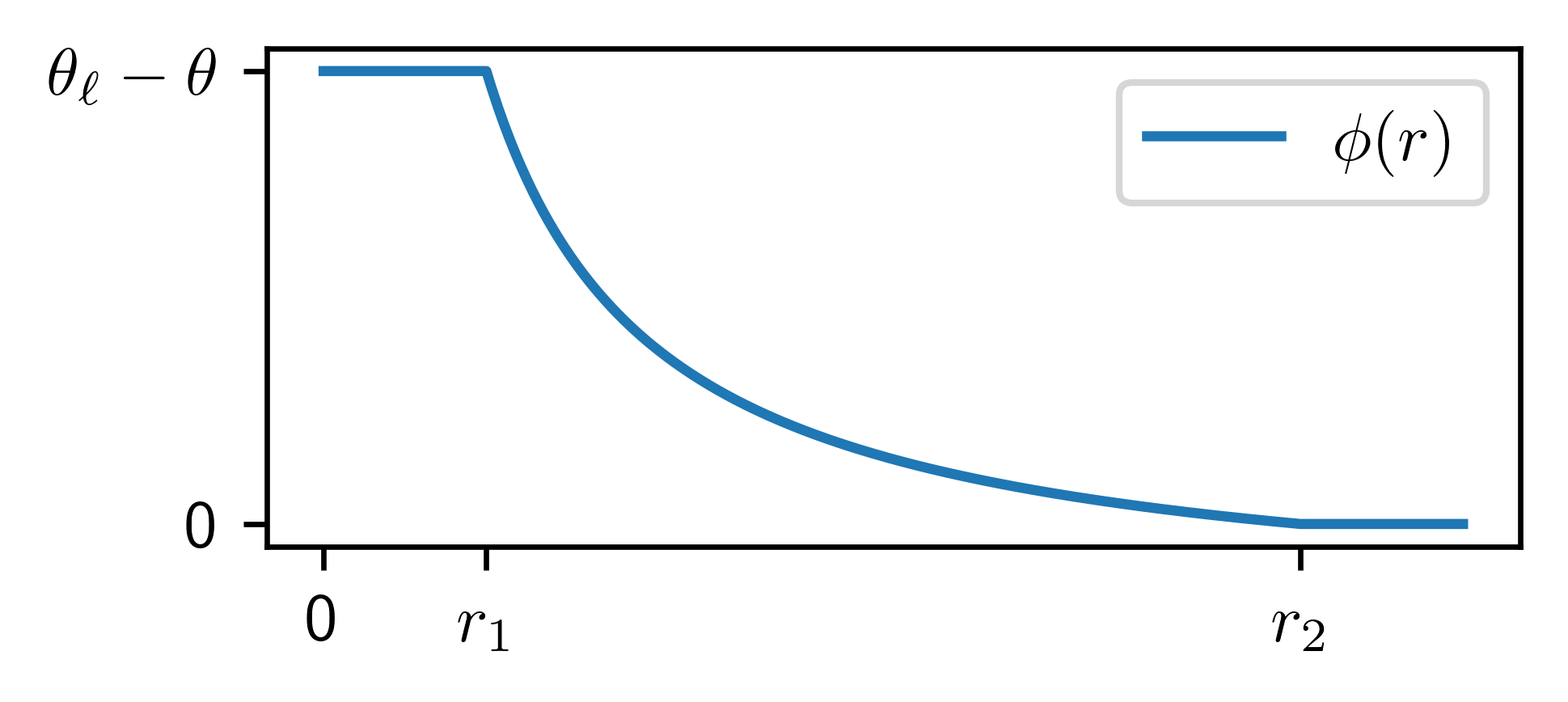}
    \hspace{-15mm}
    \includegraphics[width=.45\textwidth]{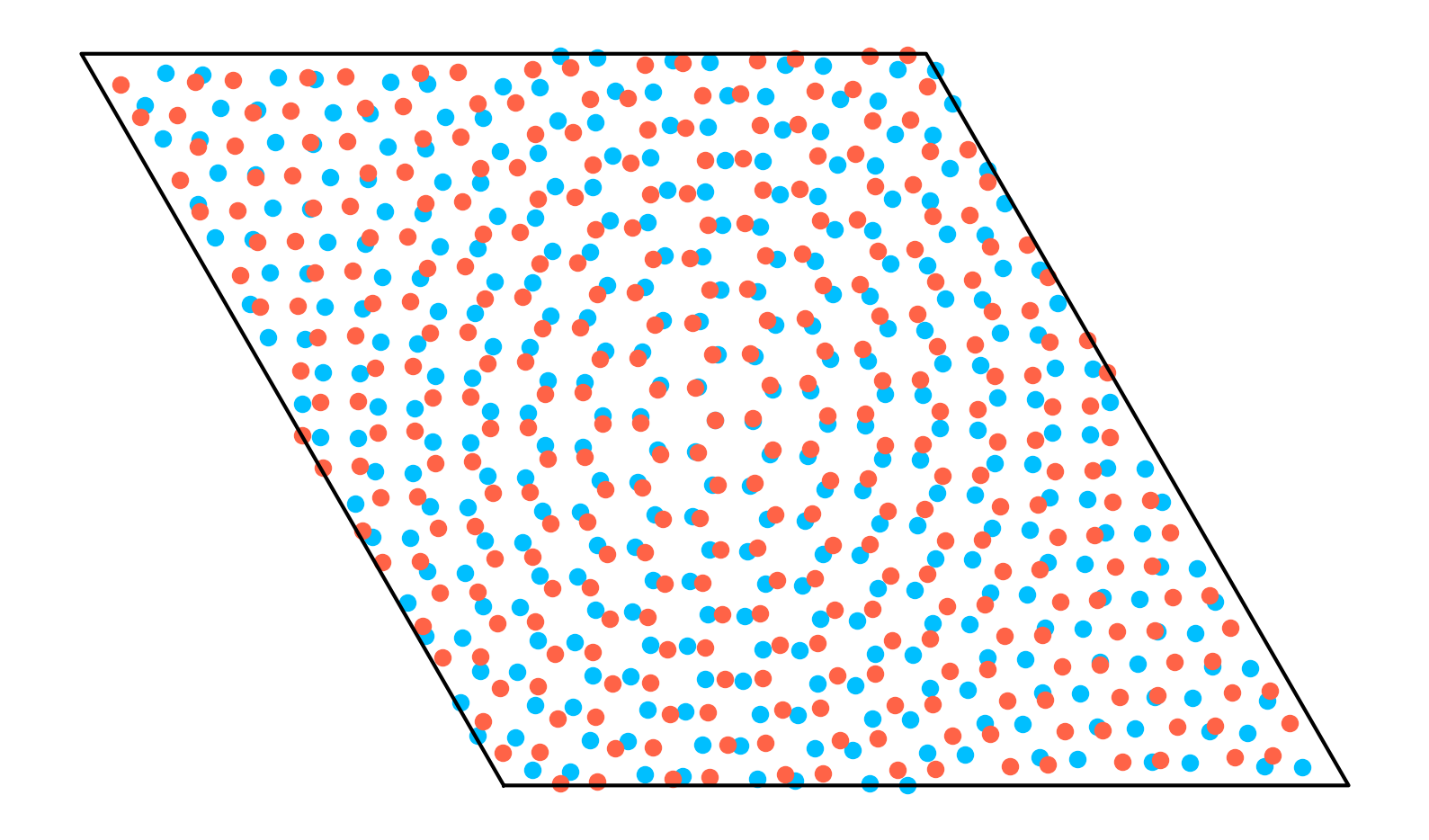}
    \hspace{-28mm}\includegraphics[width=.45\textwidth]{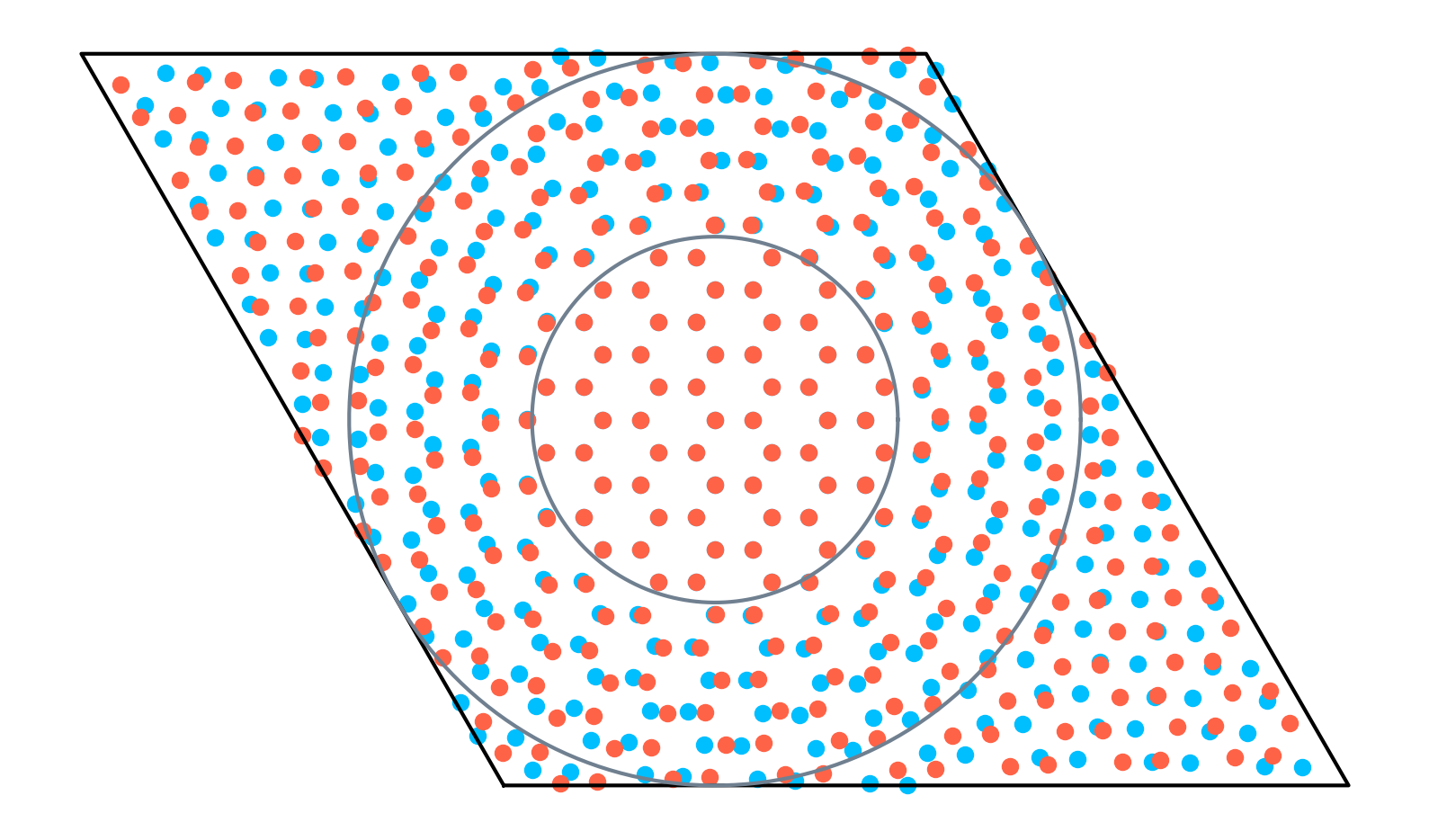}
    \caption{Left: Angle blending function $\Phi(r)$; Center: Commensurate Supercell of bilayer for $\thetasc=4.41^\circ$ (676 atoms); Right: Local twisting to $\theta=0^\circ$.}
    \label{fig:localtwist}
\end{figure}

\begin{figure}
    \centering
    \includegraphics[width=.43\textwidth]{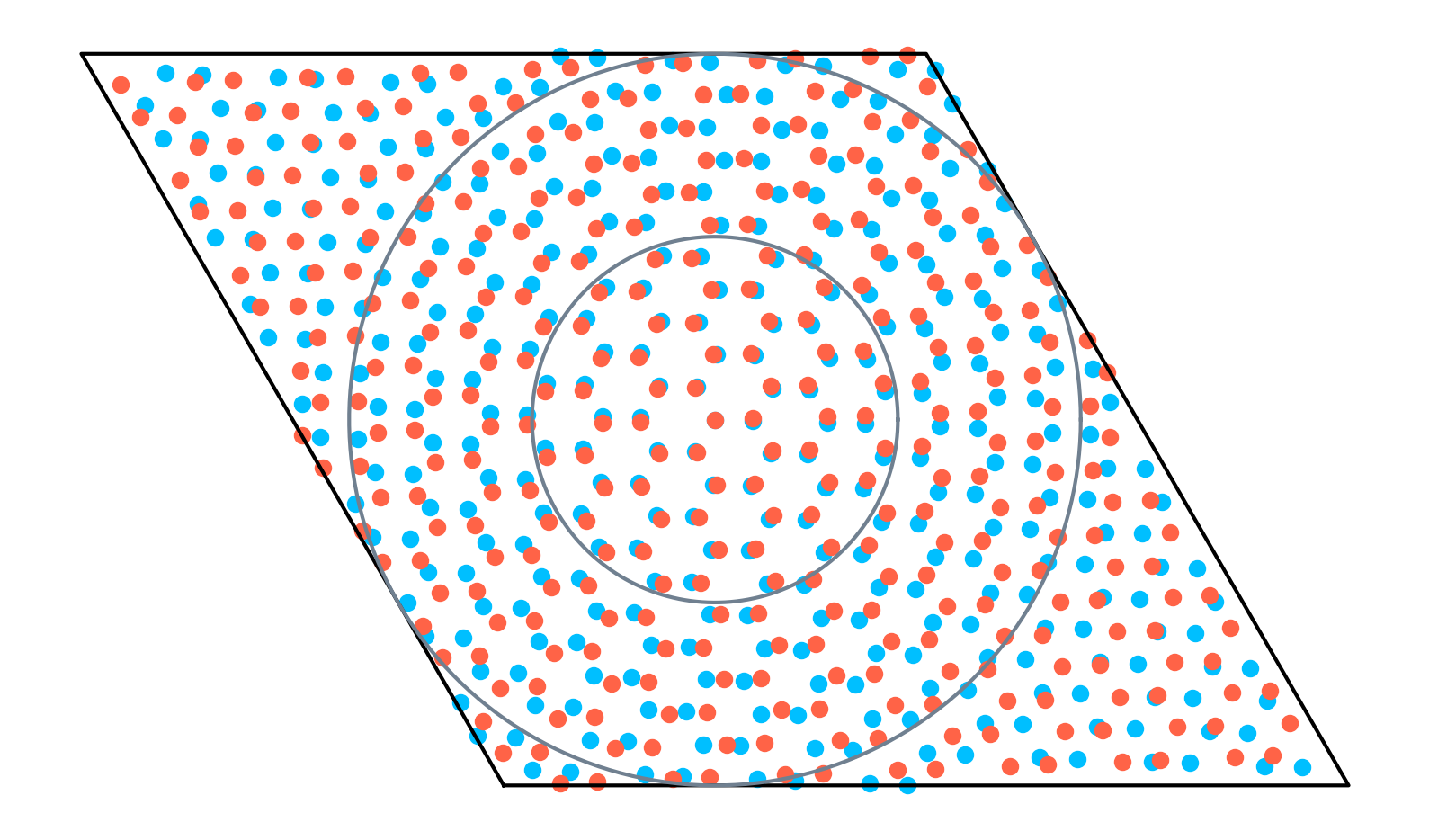}
    \hspace{-28mm}
    \includegraphics[width=.43\textwidth]{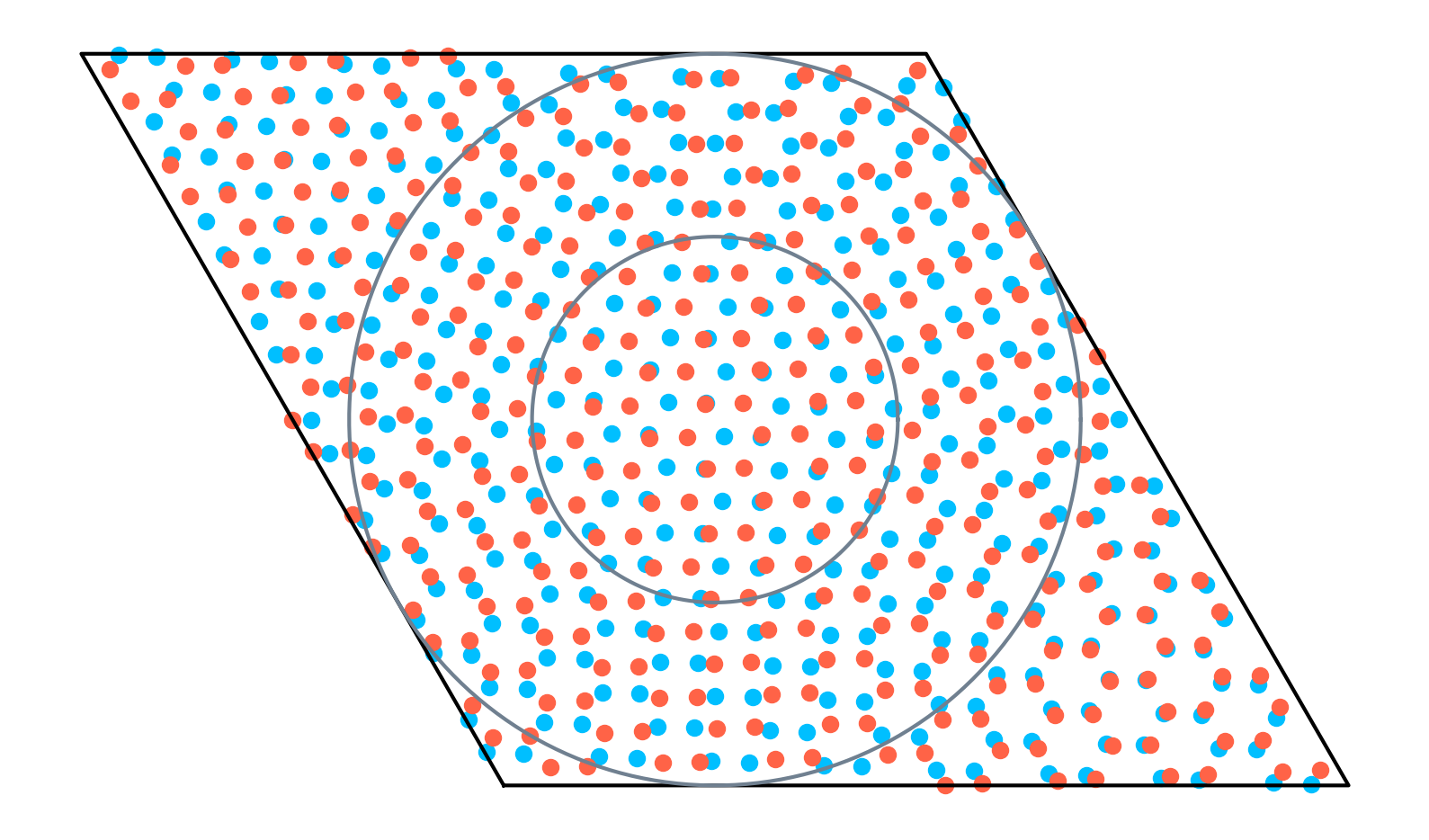}
   \hspace{-28mm}
    \includegraphics[width=.43\textwidth]{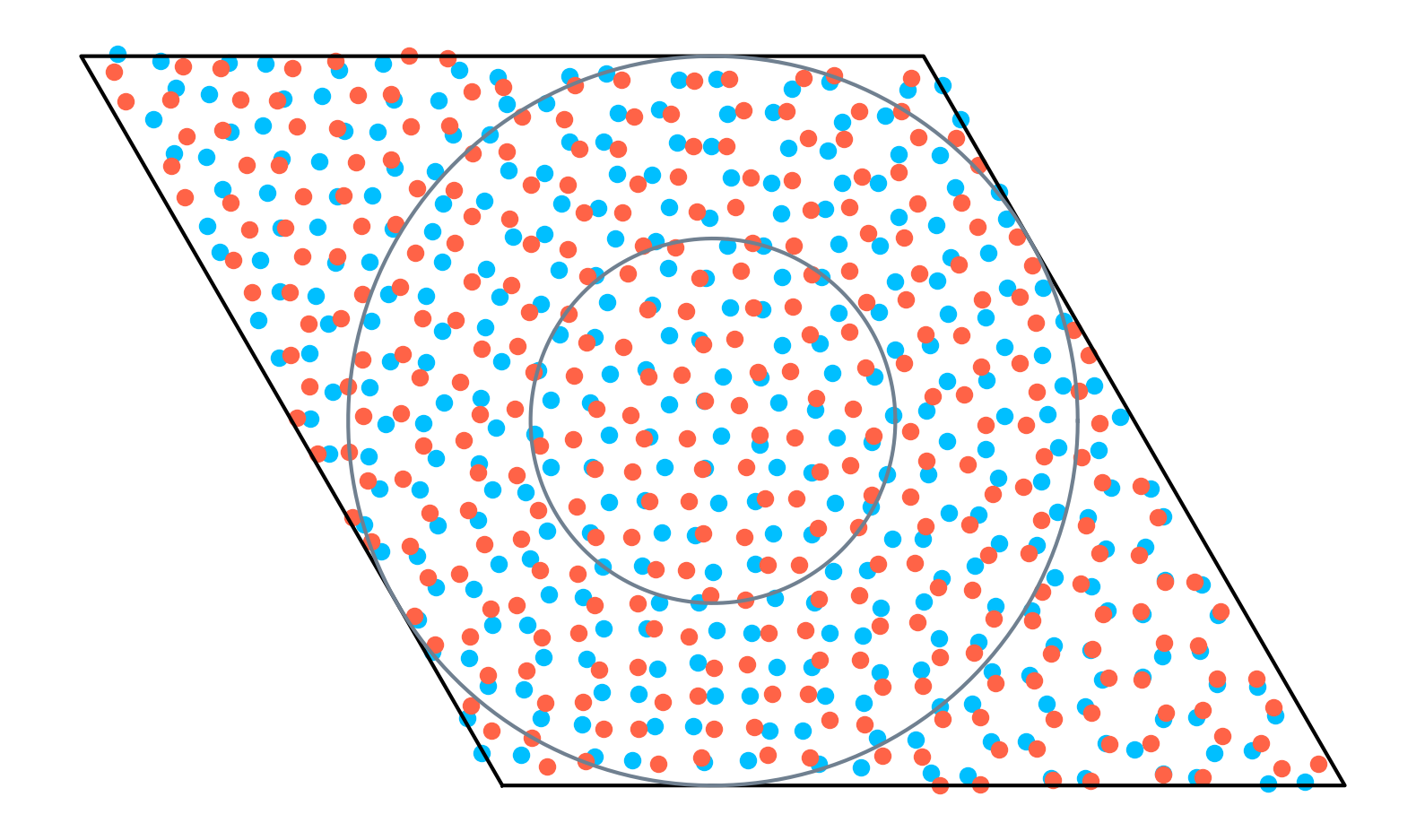}
    \caption{Left: Local twisting to $\theta=3.0^\circ$; Center: Disregistry shifting center to AB; Right: Random (normal) perturbation with stdev $0.07$ times AB distance.}
    \label{fig:disregrandom}
\end{figure}

To train beyond this limitation, we introduce a ``local twisting procedure.'' We first introduce the procedure for bilayers and then generalize it to multilayers.
%
%
Let the origin be the center of a commensurate supercell with twist angle $\thetasc$.  Pick two circles inscribed in the supercell with radii $r_1<r_2$ (see Figure~\ref{fig:localtwist}). 
%
%
Then we transform the $j^\mathrm{th}$ layer's positions by
\[\Lambda_j(x):=R\big[\Phi(|x|;(-1)^j(\theta-\thetasc)/2\big] x + b_j \]
where $R[\phi]$ is a rotation through angle $\phi$, $b_j$ is the global disregistry shift of layer $j,$ and $\Phi$ is an empirically chosen angle blending function given by 
\[\Phi(r;\phi)=\begin{cases}
        \phi, & \text{if }r<r_1, \\
        \phi\frac{r_1r_2}{r_2-r_1}\big(
        \frac{1}{r} - \frac{1}{r_2}\big), & r\in(r_1,r_2), \\
        0, & r>r_2;
\end{cases}\]
see Figure~\ref{fig:localtwist} and Figure~\ref{fig:disregrandom} for examples including twists and shifts. The rotation by $\Phi$ untwists locally by half the collective supercell angle per layer, and rotates by half the desired local twist.

Near the origin the geometry exactly corresponds to a twisted bilayer with relative shift $b=b_1-b_2$ and twist angle $\theta$, but the system globally maintains periodicity and slow variation of the displacement field. The two parameters $b, \theta$ can be chosen arbitrarily; that is, in principle this procedure is capable of generating structures that contain all possible twist angles and shifts. 

The procedure described in the foregoing paragraph can be extended to three or more alternating supercell layers. The main restriction remains the periodicity of every layer on the chosen supercell limiting the choices of global twist angles between layers. Here, we employ alternating twist angles where all adjacent layers are twisted relative to each other by alternating $\thetasc$ or $-\thetasc$, where $\thetasc$ is the commensurate twist angle corresponding to the supercell.  Let the origin again be the center of the commensurate supercell. In this case of alternating twist, we apply a similar deformation map
\[\Lambda_j(x):=R\big[\Phi(|x|;\theta_j)\big] x + b_j \]
where $b_j$ is the local shift of layer $j$ and $\theta_j$ is the relative rotation from the supercell angle, and the twist between layer $j$ and layer $j+1$ is given by $\theta_{j+1}-\theta_{j} + (-1)^j\thetasc$.


\subsubsection{Filtering the candidate dataset}

To use the uncertainty introduced in Section~\ref{sec:sub:ACE} for selecting representative configurations, we first need to construct a candidate dataset. The dataset includes monolayer, bilayer and trilayer graphene systems, with quadrilayer and pentalayer configurations reserved for out-of-distribution generalization testing (cf.~Table~\ref{tab:fittingaccuracy}). The dataset is constructed as follows:
For a single layer of graphene, we took the graphene structure discussed in Section~\ref{sec:sub:multilayerTBG} and considered $n\times n$ unit cells where $n$ ranges from 1 to 4.  For each of these configurations, we then picked an amplitude and randomly perturbed each atom's position with a normal distribution with an average of 0 and with a standard deviation equal to this amplitude.  We do this for a thousand different amplitudes evenly spaced between 0 and 0.15 times the distance between an A and B site of graphene, yielding 4000 single-layer data points to choose from.

For bilayer graphene, we make use of the local twisting method described in Section~\ref{sec:sub:sub:local}.  This method allows us to sample over arbitrary local twist angles and a selection of global twist angles (commensurate angles).  We generated datasets using global twists of $16.4^\circ$, $4.41^\circ$, and $0^\circ$ (untwisted supercell).  For the untwisted case, we used a $7\times7$ supercell.  For local twist angles, we chose 25 angles evenly distributed between 0 and $30^\circ$.  For each local angle we used the closest global twist angle as the surrounding supercell, then performed the local twist method.  We dynamically chose the disregistries used for each local twist angle to sample disregistries uniformly for each of the 25 twist angles to scale
with the size of the moir\'e cell (inversely proportional to twist angle).  The details of this process are in Appendix \ref{sec:apd:disr}.
We used a total of 479 different disregistries across 25 local twist angles.  

\par The trilayer, quadrilayer and pentalayer systems were also all generated using the local twisting method.  The commensurate twist angles used were the same as in the bilayer case (no twist, $4.41^\circ$, $16.4^\circ$).  Since any comprehensive sampling of the alternating trilayer, quadrilayer and pentalayer would yield too large of a dataset, we randomly selected disregistries, interlayer distances and local twist angle differences.  We obtained 500 unrattled and 500 rattled (randomly perturbed) samples for the trilayer, quadrilayer and pentalayer datasets. The disregistries were randomly selected uniformly over the unit cell. The interlayer distances were uniformly sampled between 2.7~\AA\ and 4~\AA\ independently for each pair of layers.  The local twist angle differences ($\theta_j$'s from the local twisting method) were uniformly chosen between $-15^\circ$ and $15^\circ$.  For the rattled samples, the rattling standard of deviation was selected uniformly from $0$ to $0.15$ times the AB distance. We could have additionally exploited the symmetries when all layers are rotated or shifted together, however, the hyperactive learning process will bias sampling away from these symmetry directions regardless.

\begin{figure}
    \centering
\includegraphics[width=1.0\linewidth]{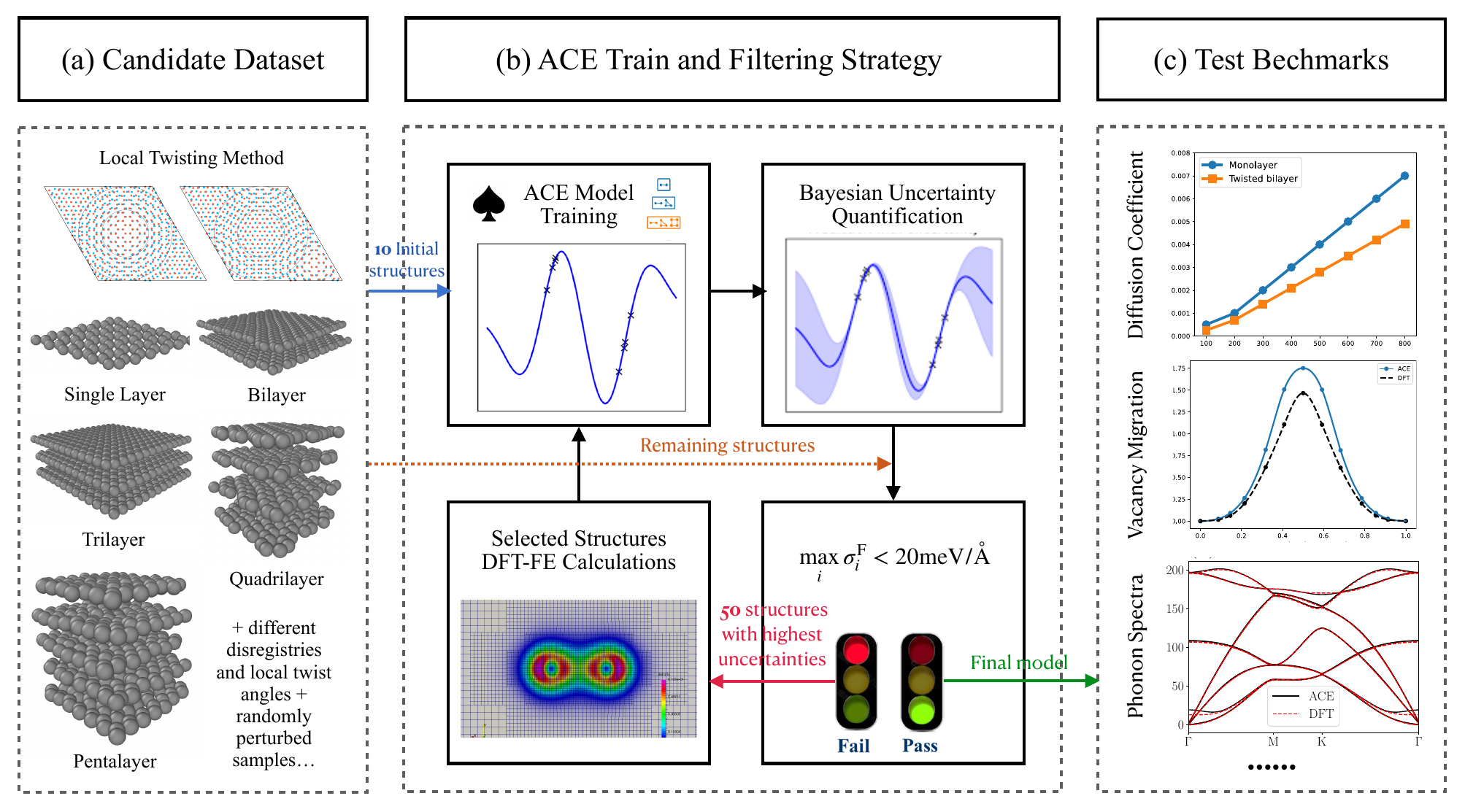}
    \caption{
{ Workflow of the proposed framework, from dataset construction and active learning filtering to ACE model training and benchmark evaluations.}
}
    \label{fig:flowchart}
\end{figure}

After constructing a large initial candidate dataset, we then design a filtering strategy using the relative force uncertainty $\sigma^{\rm F}_i$ defined in \eqref{eq:com_F} to reduce redundancy while maintaining accuracy. The correlation between $\sigma^{\rm F}_i$ and true relative force errors, as demonstrated in~\cite{hyperactive2022}, supports its effectiveness in identifying key configurations. The filtering strategy uses the following ``greedy'' approach:
\begin{enumerate}
    \setlength{\itemsep}{0pt}
    \setlength{\parskip}{2pt}
    \item Choose ACE hyperparameters: Correlation order $\nu_{\rm max} = 4$ (body-order five); Polynomial total degree 20; Cutoff radius $r_{\rm cut} = 5.5$ \AA; regression weights $w_E/w_F/w_V = 10.0/1.0/1.0$; Committee size $K = 20$. The resulting model has approximately 5k parameters.
    \item Begin with an initial dataset of 10 randomly selected configurations from the candidate set. \item Estimate the model parameters ${\bf c}$ from the current training dataset.
    \item Compute the force uncertainties $\max_i\sigma^{\rm F}_i$ for all remaining candidate training structures. Add the 50 configurations with the highest force uncertainties to the training set.
    \item Repeat from Step 3 until the maximum force uncertainty, $\max_i \sigma_i^{\rm F}$, is reduced to a specified threshold of 20~meV/\AA.
\end{enumerate}
This filtering strategy effectively reduced the dataset to approximately 5\% of the original candidates while achieving strong generalization between training and testing errors, as we demonstrate in the next section. The workflow is illustrated in Figure~\ref{fig:flowchart}.

\subsubsection{Fine-tuning}
It is all but impossible to fit an MLIP that has good quantitative accuracy across all possible simulation tasks for a given material. The models we report will be shown to have excellent accuracy in-distribution (near the training set) and behave robustly with good qualitative accuracy out-of-distribution. In order to obtain quantitatively accurate result on new simulation tasks, it may be necessary to extend the training data in a way that is representative of the new tasks, or within an active learning procedure. Given the low cost of fitting a linear ACE potential it is straightforward to then simply refit the model to the expanded dataset, normally with the same hyperparameters.  We will demonstrate this on a limited set of out-of-distribution tests. We call this procedure ``fine-tuning'' to indicate the analogy with fine-tuning foundation models. 

\section{Results}
\label{sec:results}
We evaluate the robustness and transferability of the fitted ACE potential trained using the methods introduced in the previous section. We conduct comprehensive numerical tests on twisted graphene at various angles and layer configurations. Section~\ref{sec:sub:fitting} assesses the force accuracy of the ACE model. Section~\ref{sec:sub:minimization} examines the geometry optimization of graphene configurations to evaluate their static mechanical properties. Section~\ref{sec:sub:md} presents the performance of the ACE potential in molecular dynamics simulations, demonstrating its capability to compute both static and dynamic properties of twisted bilayer graphene. Section~\ref{sec:sub:phonon} investigates the accuracy of phonon spectra. 

To compare and demonstrate the effectiveness of training on locally twisted training structures as well as our dataset filtering strategies, we trained various potentials using different dataset constructions. Table~\ref{tbl:models} summarizes the potentials used in this study and specifies the scenarios in which they have been applied.

\begin{table}
{\renewcommand{\arraystretch}{1.3}
    \begin{center}
\begin{tabular}{p{2cm}|p{9cm}|p{5.8cm}}
\toprule
\textbf{Label} & \textbf{Description} & \textbf{Application} \\ 
\midrule
\ACEstar & Fitted to dataset as described in Section~II~C, (0.5k structures, 154k atoms) & Throughout the paper \\ 
\midrule
\quad \newline 
ACE (gl) \newline ACE (lt) \newline ACE (gt)
& Randomly selected configurations from candidate dataset.\newline
- Globally and locally twisted (gl), (0.3k struct., 225k atoms) \newline
- Locally twisted (lt), (0.3k struct., 203k atoms) \newline
- Globally twisted (gt), (0.3k structures, 240k atoms) 
& Section~III~B: Geometry optimization \\ 
\midrule
%
\quad \newline ACE (mo) \newline ACE (bi)
& Randomly selected configurations from candidate dataset: \newline
- Only monolayer (mo), (0.9k structures, 64k atoms) \newline
- Only bilayer (bi), (0.4k structures, 270k atoms) 
& Section~III~A: Fitting accuracy
\newline Section~III~C: Molecular Dynamics\\ 
\midrule
\quad \newline KC \newline pacemaker 
& Comparison interatomic potential models: \newline
- Kolmogorov–Crespi potential~\cite{KolmogorovCrespi2005} \newline
- Nonlinear ACE potential for general carbon systems~\cite{2022_drautz_ace_C}, \newline (17k structures, 366k atoms)
& Section~III~A: Fitting accuracy; 
\newline Section~III~B: Geometry optimization; 
\newline Section~III~D: Phonon spectra. \\ 
\bottomrule
\end{tabular}
\end{center}
}
\caption{List of potentials tests throughout Section~\ref{sec:results}, including several variants of our linear ACE potential and two comparison potentials from the literature. The reported number of structures and total number of atoms across all structures specifies the amount of data from which the models are trained.
\label{tbl:models}
}
\end{table}

\subsection{Fitting Accuracy}
\label{sec:sub:fitting}
\begin{figure}[t!] 
\centering
\includegraphics[height=12cm]{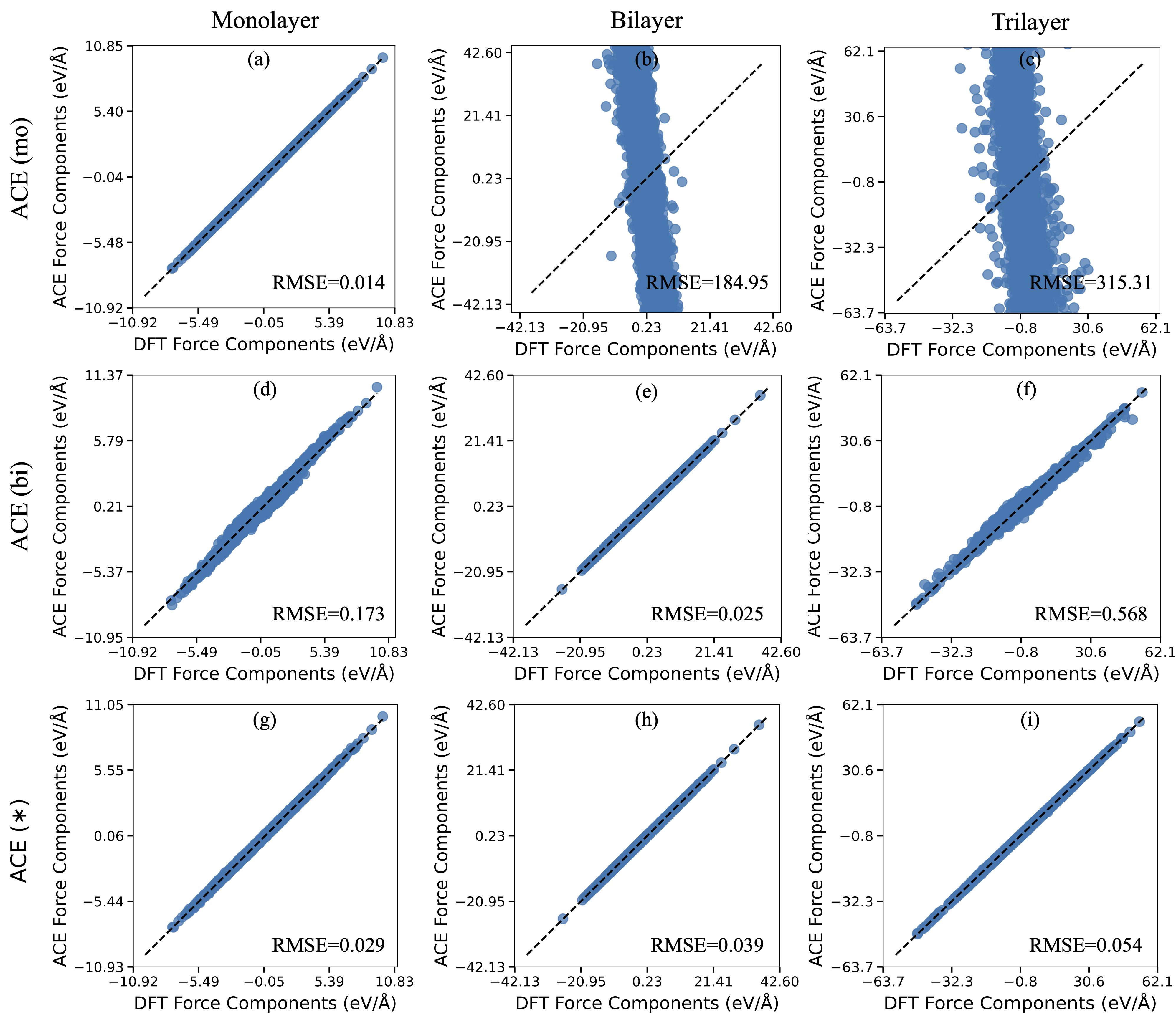}
\caption{Test accuracy for three ACE potentials: (a-c) for ACE (mo) trained only on mono-layer data, (d-f) for ACE (bi) trained on only bi-layer data and (g-i) for \ACEstar\ trained on the full dataset described in \S~\ref{sec:sub:data}; cf. Table~\ref{tbl:models} for further specifications of the three models.
}
\label{fig:bilayer_fitting}
\end{figure}

We investigate the accuracy and the generalization of several fitted ACE models. We compare the performance of ACE models for multilayer twisted graphene systems, evaluating two scenarios: one where the model was trained on twisted bilayer configurations and another where it was trained on monolayer configurations. After developing these two models, we tested their force-fitting accuracy on monolayer, twisted bilayer, and twisted trilayer configurations separately. The results are summarized in Figure~\ref{fig:bilayer_fitting} and in Table~\ref{tab:fittingaccuracy}. The Kolmogorov-Crespi potential has fairly poor accuracy across all tests. The generic pacemaker C potential performs surprisingly well given it is trained primarily on bulk data, only being outperformed in accuracy by the \ACEstar \ potential trained on mono-, bi- and trilayer structures. We will observe similar behaviour in many of the remaining tests. 

Figure~\ref{fig:bilayer_fitting} provides a more detailed description of the force accuracy of the ACE potentials. In each column, the three subplots represent the predicted forces versus the DFT reference forces for monolayer, twisted bilayer, and twisted trilayer configurations, respectively. 
%
%
The results demonstrate that training only on mono-layer configurations (ACE (mo)) does not yield any acceptable generalization to bi- and tri-layers. This is unsurprising since the cutoff radius for the ACE potentials is larger than the interlayer distance. Training on only bilayer configurations results in moderately accurate generalization to both monolayer and twisted trilayer configurations. Note in particular that the generic {\tt pacemaker} model~\cite{2022_drautz_ace_C} has far better accuracy on all multi-layers except bilayers. {
Our approach benefits from a training dataset specifically curated for twisted graphene systems such as local twist angle variations and stacking disorder. This task-specific dataset results in higher accuracy of the fitted linear ACE model on our specific tests. This accuracy improvement is likely independent of the chosen MLIP model. 
%
%
} 

The final potential trained on the filtered dataset, \ACEstar, provides significantly better accuracy across all multi-layers, validating both the accuracy and the generalization ability of the potential trained using the filtering strategy. Moreover, the results demonstrate that an \ACEstar\ model trained on single, bi- and tri-layer configurations provides excellent accuracy also on quadrilayer and pentalayer configurations. This is likely due to a very short interaction distance in multi-layer graphene and it is unclear whether such a result would generalize to more complex materials.

\begin{table}
\caption{\label{tab:fittingaccuracy} Test accuracy in terms of RMSE (relative to DFT) for the three ACE potentials trained in this work, the pacemaker, and the KC potentials. All values are reported in~eV/\AA. Entries with an RMSE below 0.1~eV/\AA{} are highlighted.  } 
{\renewcommand{\arraystretch}{1.3}
\begin{ruledtabular}
\begin{tabular}{lccccc}
 & Monolayer & Bilayer & Trilayer & quadrilayer & Pentalayer \\
\hline
KC & 0.641 & 1.207 & 4.368 & 4.782 & 4.869 \\
pacemaker & 0.106 & 0.126 & 0.441 & 0.454 & 0.468 \\
ACE (mo) & {\bf 0.014} & 184.952 & 315.314 & 316.468 & 306.203 \\
ACE (bi) & 0.173 & {\bf 0.025} & 0.568 & 1.149 & 1.502 \\
\ACEstar & {\bf 0.029} & {\bf 0.039} & {\bf 0.054} & {\bf 0.089} & {\bf 0.093} \\
\end{tabular}
\end{ruledtabular}
}
\end{table}

\subsection{Geometry Optimization}
\label{sec:sub:minimization}
We test the ability of the fitted ACE potentials to accurately predict equilibrium crystalline and defective structures.
We perform geometry optimization experiments for various configurations of twisted multilayer graphene, including defect formation, defect migration, and stress-strain curves. 

\subsubsection{Stability of structure equilibration}
\label{sec:sub:sub:structure_min}
To facilitate comparison and to test the performance of the filtering strategy introduced in Section~\ref{sec:sub:data}, we trained several  ACE models using different dataset combinations. The model trained using the filtering strategy described in Section~\ref{sec:sub:data} is denoted by \ACEstar. We compare against ACE models trained on a variety of randomly selected globally and locally twisted datasets; see Table~\ref{tbl:models} for the details.

The systems for which we conduct geometry optimization include twisted bilayer graphene with a twist angle of $4.41^\circ$, both pure and with point defects, including vacancies, Stone-Wales (SW) defects~\cite{ma2009stone}, intralayer interstitials and interlayer interstitials. The geometry minimization solver used is \textsf{L-BFGS}~\cite{shajan2023geometry}, with a force tolerance set to $10^{-3}$eV/\AA. Our results show that only models trained on both globally and locally twisted configurations perform reliably across all tests. 

Table~\ref{tab:min} summarizes geometry optimization results for all models and all configurations. A check mark indicates that the minimization procedure successfully converged to a local minimum, while a cross mark signifies that the procedure either failed or did not converge within the 1000 iterations. 
An unsuccessful optimisation typically entails energies descreasing into the negative thousands of eV range and forces reaching O($10^5$) eV/\AA~range.
The models trained using the filtering strategy and those trained on locally twisted configurations, as introduced in Section~\ref{sec:sub:data}, performed well, consistently achieving equilibrium structures for all defective cases. Models trained solely on globally twisted configurations failed to converge for the interlayer and intralayer interstitial cases, likely due to the stronger interactions in those cases. The KC and pacemaker potentials also converged successfully for all cases.

Table~\ref{tab:min_angles} presents analogous results of geometry relaxation for five commensurate twisted bilayer graphene configurations with twist angles ranging from 4.41$^\circ$ to 33.65$^\circ$. Models trained using the filtering strategy and those trained on locally twisted configurations (introduced in Section~\ref{sec:sub:data}) performed well. 
%
%
In contrast, models trained exclusively on globally twisted configurations failed to converge for the largest twist angle of 33.65$^\circ$, likely due to extrapolation challenges.

The number of optimization steps shown is brackets are an indicator of the smoothness of the potential energy surface, facilitating comparisons across different interatomic potentials (e.g., ACE, KC, pacemaker). By this measure, all ACE models using locally twisted training structures exhibits superior convergence behavior, suggesting a qualitatively smoother energy landscape.


\begin{table}
\caption{\label{tab:min} Validation of minimization for homogeneous and defected configurations in twisted bilayer systems with a twist angle of 4.41$^\circ$. Here, a check mark~(\checkmark) indicates successful geometric relaxation, with the number in parentheses denoting the number of iterations required for convergence.  A cross~($\times$) indicates that the relaxation failed to converge. A large number of iterations or convergence failure indicate a non-smooth or even unphysical potential energy surface.
}
{\renewcommand{\arraystretch}{1.3}
\begin{ruledtabular}
\begin{tabular}{cccccc}
 & Homogeneous & Vacancy & Stone-Wales & Interlayer interstitial & Intralayer interstitial \\
\hline
KC & $\checkmark (55)$ & $\checkmark (65)$ & $\checkmark (719)$ & $\checkmark$(260) &  $\checkmark$(448) \\
pacemaker & $\checkmark (47)$ &  $\checkmark (505)$ & $\checkmark (736)$ & $\checkmark (353)$ & $\checkmark (550)$ \\
ACE (gt) & $\checkmark (71)$ & $\checkmark (132)$ & $\times$ & $\times$ & $\times$ \\
ACE (lt) & $\checkmark (48)$ & $\checkmark (257)$ & $\checkmark (394)$ & $\checkmark (257)$ & $\checkmark (360)$ \\
ACE (gl) & $\checkmark (47)$ & $\checkmark (224)$ & $\checkmark (309)$ & $\checkmark (253)$ & $\checkmark (298)$ \\
\ACEstar & $\checkmark (48)$ & $\checkmark (235)$ & $\checkmark (290)$ & $\checkmark (242)$ & $\checkmark (340)$
\end{tabular}
\end{ruledtabular}
}
\end{table}

\begin{table}
\caption{\label{tab:min_angles} Validation of minimization for commensurate twisted bilayer configurations with various twist angles.}
{\renewcommand{\arraystretch}{1.3}
\begin{ruledtabular}
\begin{tabular}{ccccccc}
Twisted angles & $4.41^\circ$ & $16.43^\circ$ & $24.21^\circ$ & $29.64^\circ$ & $33.65^\circ$ \\
Number of atoms & 676 & 196 & 3844 & 7396 & 1444 \\ \hline
KC & $\checkmark$ (49) & $\checkmark$ (26) & $\checkmark$ (27) & $\checkmark$ (28) & $\checkmark$ (26) \\
pacemaker & $\checkmark$ (43) & $\checkmark$ (26) & $\checkmark$ (23) & $\checkmark$ (26) & $\checkmark$ (24) \\
ACE (gt) & $\checkmark$ (38) & $\checkmark$ (32) & $\checkmark$ (27) & $\times$ & $\times$ \\
ACE (lt) & $\checkmark$ (37) & $\checkmark$ (28) & $\checkmark$ (31) & $\checkmark$ (33) & $\times$ \\
ACE (gl) & $\checkmark$ (35) & $\checkmark$ (31) & $\checkmark$ (27) & $\checkmark$ (34) & $\checkmark$ (39) \\
\ACEstar & $\checkmark$ (36) & $\checkmark$ (30) & $\checkmark$ (29) & $\checkmark$ (32) & $\checkmark$ (41) \\
\end{tabular}
\end{ruledtabular}
}
\end{table}


\subsubsection{Defect formation energies}
\label{sec:sub:sub:structure_min_defect}
The previous section demonstrates the stability of geometry optimization. Next, we test quantitative accuracy of the predicted minimizer by comparing predicted defect formation energies against their corresponding DFT values. These tests assess out-of-distribution performance, since the training set does not contain defective structures. The defect-free system is a $4.41^\circ$ degree twisted bilayer structure containing 676 atoms. Interatomic forces are relaxed in all the systems using the trained interatomic potentials. The reference DFT energetics are computed on the relaxed structures using the trained potentials. Further relaxation of DFT interatomic forces is not conducted.
%
%
Given that the model ACE (gt) trained only on globally twisted configurations does not perform well, we limit these tests to the ACE (lt), ACE (gl) and \ACEstar\ models. The results are summarized in Table~\ref{tab:formation}. 
We observe that the \ACEstar\ model (generated using the active learning strategy) provides significantly higher accuracy on all tests, with the exception of a single test where the pacemaker potential achieves slightly better accuracy. The models trained solely on locally twisted configurations and those trained on combined data all show relatively crude accuracy, comparable with the general purpose pacemaker potential. 
{
To improve model transferability to defective configurations, we additionally included 50 near-equilibrium defect structures (vacancy, Stone–Wales, and interstitial types) in the training set. The resulting ACE (ft) shows consistently improved accuracy in predicting defect formation energies compared to models trained without such data.
}

\begin{table}
\caption{\label{tab:formation}
Point defect formation energies in bilayer graphene with a twist angle of $4.41^\circ$ for three ACE potentials. The relative error is given in the bracket. Entries within $\pm5\%$ of the reference values and the best-performing results across all methods are highlighted.}
{\renewcommand{\arraystretch}{1.3}
\begin{ruledtabular}
\begin{tabular}{ccccc}
 & Vacancy & Stone-Wales & Interlayer interstitial & Intralayer interstitial \\
\hline
DFT & 8.419 & 4.948 & 9.017 & 7.883 \\
\hline
KC & 7.707 (-8.46\%) & 5.430 (9.74\%) & 12.463 (38.22\%) & 7.183 (-8.88\%) \\
pacemaker & 7.053 (-16.23\%) & \textbf{4.960 (-0.24\%)} & 7.867 (-12.75\%) & 6.797 (-13.78\%) \\
ACE (lt) & 7.186 (-14.64\%) & 5.710 (15.41\%) & 7.221 (-19.91\%) & 6.408 (-18.70\%) \\
ACE (gl) & 7.825 (-7.06\%) & 5.540 (11.97\%) & 7.885 (-12.55\%) & 6.971 (-11.57\%) \\
\ACEstar & \textbf{8.103 (-3.75\%)} & \textbf{5.144 (3.96\%)} & 8.256 (-8.44\%) & \textbf{8.031 (1.88\%)} \\
{ ACE (ft)} & { \textbf{8.265 (-1.83\%)}} & { \textbf{5.096 (2.99\%)}} & { \textbf{8.393 (-6.92\%)}} & { \textbf{7.987 (1.29\%)}}
\end{tabular}
\end{ruledtabular}
}
\end{table}

\subsubsection{Equilibration of twisted trilayer graphene}
\label{sec:sub:sub:min-check}
\begin{figure}
\centering
\normalsize ACE (gt) \hspace{6cm} ACE (lt) \\[1mm]
\includegraphics[trim={0 8cm 0 2cm}, clip, width=0.9\linewidth]{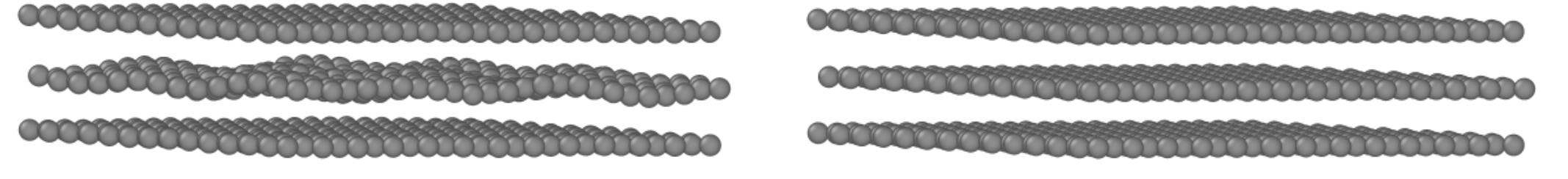}
\caption{Equilibrium structure of alternately twisted $4.41^\circ$ trilayer graphene, minimized using the ACE (gt) potential trained on globally twisted configurations ({\bf left}) and the ACE (lt) potential trained on locally twisted configurations ({\bf right}). The unphysical out-of-plane displacement in the left-hand figure are approximately 1.1\AA.}
\label{fig:locally-twisted}
\end{figure}

We perform geometry optimization of trilayer graphene with alternating twist angle of $\pm 4.41^\circ$. This test is restricted to the ACE (gt) and ACE (lt) potentials. 
%
Figure~\ref{fig:locally-twisted} visualizes the two computed equilibrium structures.
The potential ACE (lt) trained on locally twisted configurations predicts a realistic equilibrium structure. By contrast, the potential ACE (gt) trained only on globally twisted configurations predicts an equilibrium with noticeable undulations and distortions in the layers, indicating significant interlayer interactions that are incorrectly predicted. The \ACEstar\ model shows similar performance to the ACE (lt) model.

\vspace{0.5cm}

\subsubsection{Vacancy migration}
\label{sec:sub:sub:vac-migration}
{ As a more stringent out-of-distribution test we now consider the prediction of a vacancy migration pathway, calculated using the Nudged Elastic Band (NEB) method~\cite{jonsson1998nudged}. We consider a single vacancy in the upper layer of bilayer graphene with a twist angle of $4.41^\circ$.
This test is restricted to the \ACEstar\ model trained using the filtering approach, and to the pacemaker and KC comparison potentials.} This is an out-of-distribution test: the training sets for neither MLIPs employed in this test contain any defect structures, and in particular do not contain structures close to the target migration pathway.

\begin{figure}[ht!]
        \centering
        \includegraphics[height=7cm]{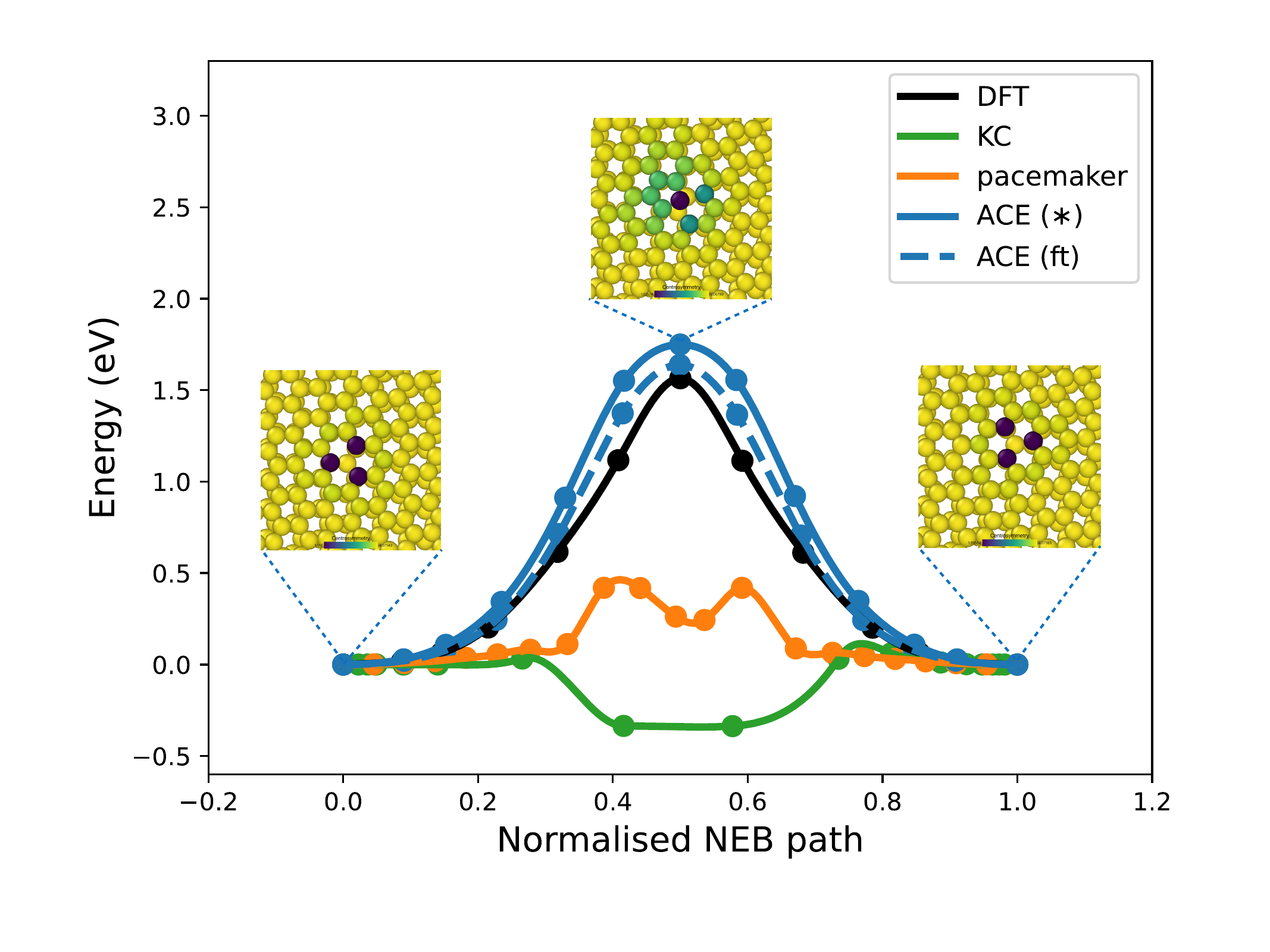}~~
    \caption{The NEB path for a vacancy migration in twisted bilayer graphene (cf. Sec.~\ref{sec:sub:sub:vac-migration}), migration energy barrier and the defect core structures of initial, saddle and final states along the NEB path. Only the \ACEstar\ and its fine-tuned potential, ACE (ft), provide a qualitatively correct prediction. }
    \label{fig:migration}
\end{figure}

The results are reported in Figure~\ref{fig:migration}. While both the pacemaker and KC models fail to predict qualitatively reasonable migration profiles, the \ACEstar\ potential predicts an energy path and transition energy barrier that are qualitatively correct and only have a moderately small error when compared against the DFT predictions.  
To assess the impact of fine-tuning, we include the DFT reference configurations along the NEB path in the training set and retrain the model. The fine-tuned model, ACE (ft), shows improved accuracy and in particular reproduces the DFT energy barrier.

\subsubsection{Stress-strain curves}
\label{sec:sub:sub:stress-strain}
Stress-strain curves offer essential insight into the mechanical behavior of materials, particularly their elastic properties and failure mechanisms. In multilayer graphene, this response is highly sensitive to stacking order, interlayer interactions, and structural defects. In this section, we evaluate the stress-strain behavior of several multilayer graphene systems under uniaxial tension to assess the mechanical trends and the predictive capability of the fitted \ACEstar\ potential. Since our potentials were not trained on strained systems, this is a (mild) out-of-distribution test, hence we cannot expect high accuracy.

Starting from fully relaxed atomic structures, with both atomic positions and cell parameters optimized, we apply uniaxial strain along the armchair direction by incrementally increasing the deformation up to 20\%. At each strain level, the structure is re-relaxed using the L-BFGS algorithm~\cite{liu1989limited} with a force convergence threshold of 0.001eV/\AA, while keeping the cell fixed along the transverse directions. The virial stress tensor is then computed to extract the longitudinal stress component.

\begin{figure}
\centering 
\includegraphics[height=5cm]{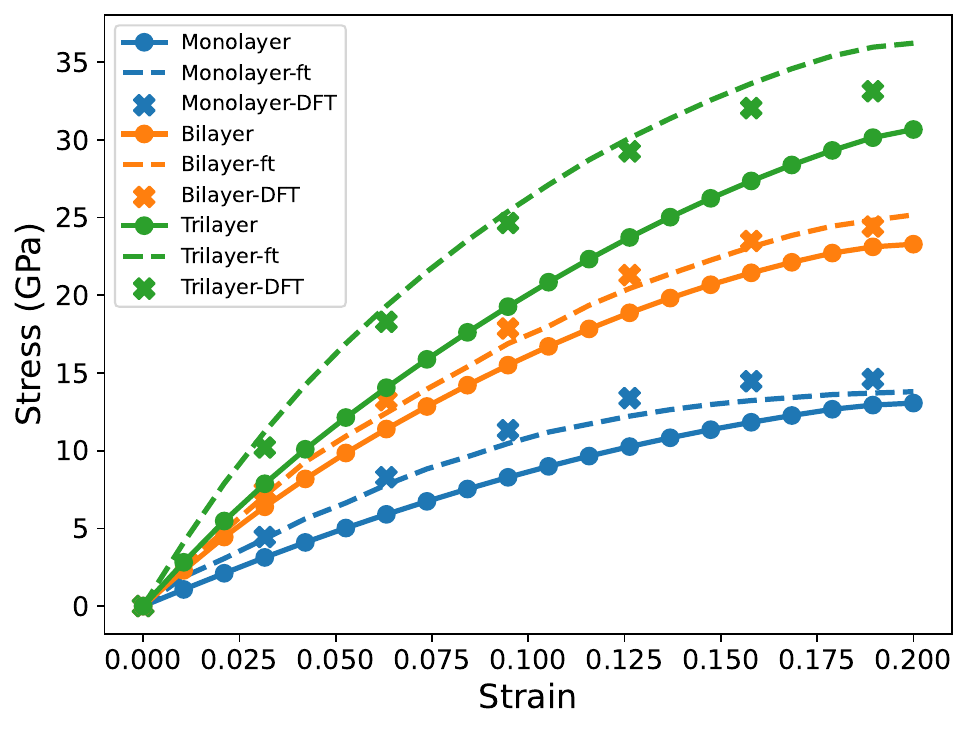}~~
\includegraphics[height=5cm]{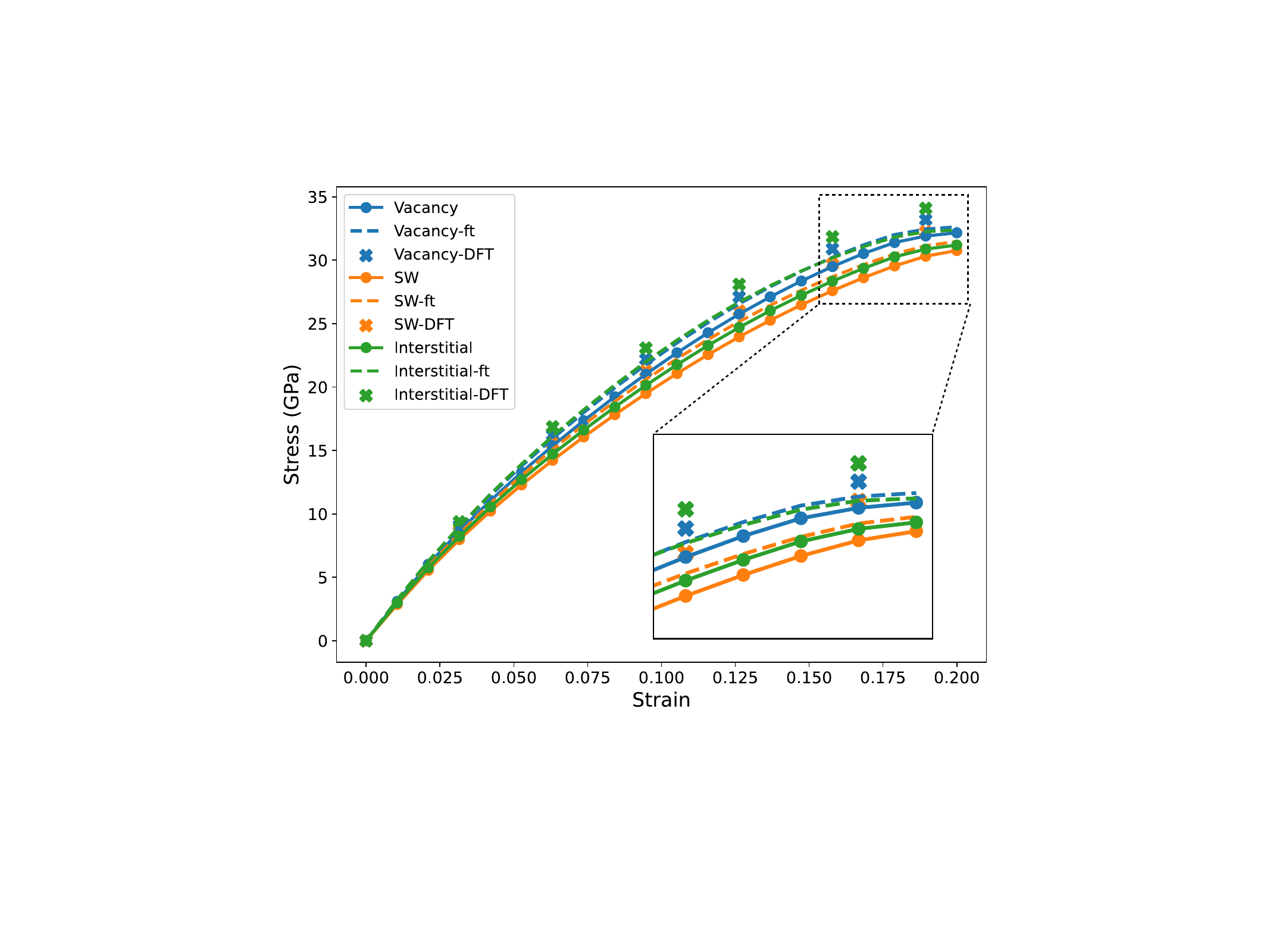}
\caption{Stress-strain curves computed using the \ACEstar\ potential and the fine-tuned ACE potentials. {\bf Left:} Monolayer, bilayer (with a twist angle of $4.41^\circ$), and trilayer (with alternating twist angles of $\pm 4.41^\circ$) graphene under uniaxial tension along the armchair direction. {\bf Right:} Trilayer graphene systems containing different types of point defects, including a single vacancy, a Stone-Wales (SW) defect, and an interlayer interstitial. }
\label{fig:stress-strain}
\end{figure}

We examine several representative systems, including monolayer graphene, bilayer graphene with a twist angle of $4.41^\circ$, and trilayer graphene with alternating twist angles of $\pm 4.41^\circ$. Additionally, we consider bilayer configurations with point defects such as vacancies, Stone–Wales defects, and interlayer interstitials. The resulting stress-strain curves are shown in Figure~\ref{fig:stress-strain}. In the linear elastic regime, the slope of each curve yields the Young’s modulus, $E = \sigma / \epsilon$, characterizing the stiffness of the material. Among the systems studied, the trilayer structure exhibits the highest Young’s modulus, indicating increased stiffness with additional layers. The trends are consistent with DFT reference calculations. { We also examine the effect of fine-tuning by incorporating the DFT reference strained configurations (60 configurations) into the training set and retraining the model.} The resulting fine-tuned model exhibits clear improved agreement with the DFT reference across both test cases. 

\subsection{Molecular Dynamics}
\label{sec:sub:md}

After systematically examining the accuracy and stability of the fitted ACE models in geometry optimization tasks, we now conduct similar tests of the models' performance in molecular dynamics (MD) simulations. { All molecular dynamics simulations performed in this section were conducted using a Langevin thermostat, targeting the desired temperature with a friction coefficient of 0.01 $\textrm{fs}^{-1}$ and 1fs time-step. As in the previous section, we use twisted bilayer configurations with a twist angle of $4.41^\circ$ with preiodic boundary conditions throughout this section unless otherwise specified.} { All MD simulations were performed using the Atomic Simulation Environment (ASE)~\cite{ase2002}.}

\subsubsection{MD Stability}
\label{sec:sub:sub:md-local-check}
{ We perform molecular dynamics simulations of $4.41^\circ$ twisted bilayer graphene at 1000K, using the ACE (gt) and ACE (lt) potentials.} The results are shown in Figure~\ref{fig:md} { in Appendix~\ref{sec:apd:numerics}.} 
The MD simulation using ACE (gt) (trained only on globally twisted configurations) becomes unstable after around { 2500 time steps}. After this point, the temperature rises sharply, accompanied by significant changes in the potential energy and pressure. This confirms our previous observations that the potential energy surface generated by the ACE (gt) model, trained only on globally twisted configurations, contains ``unstable regions'' that lead to non-physical behavior in the simulations. By contrast, the MD simulation using ACE (lt) (trained on locally twisted configurations) remains stable for the entire trajectory of { 20,000 steps}, further supporting the value of using locally twisted configurations during model training.

\begin{figure}
\centering
\includegraphics[height=10cm]{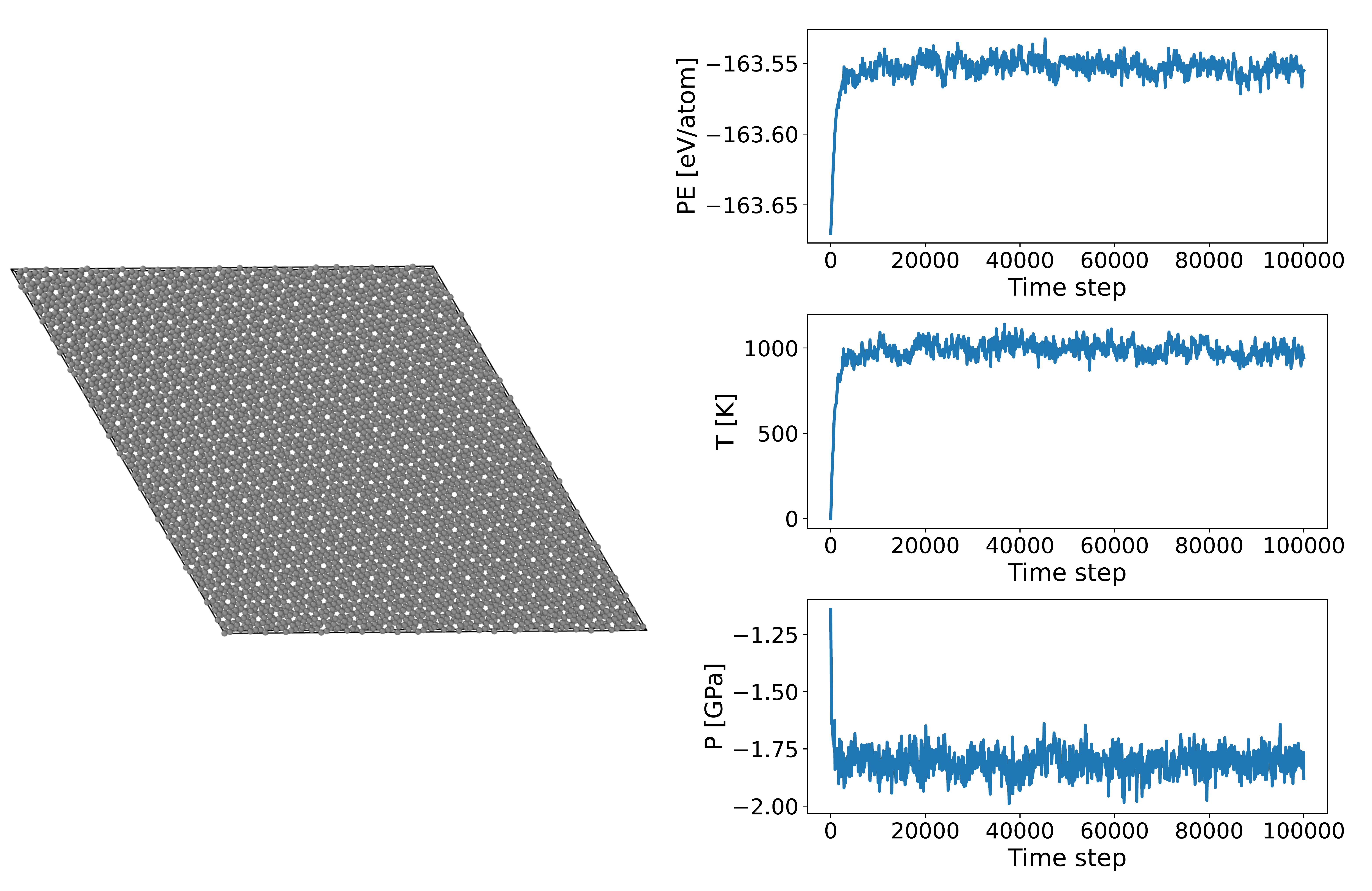}
\caption{{\bf Left:} Atomic configuration of a commensurate unit cell of bilayer graphene twisted by $29.64^\circ$, featuring an interlayer interstitial defect, containing a total of 7397 atoms. {\bf Right:} Evolution of thermodynamic properties at 1000K over $10^5$ molecular dynamics steps. The {\ACEstar~}exhibits stable molecular dynamics for the entire trajectory.}
\label{fig:geo}
\end{figure}

\label{sec:sub:sub:large-scale-md}
Next, we demonstrate the stability of MD for bilayer graphene in a large domain and much longer simulation time, only for the \ACEstar\ potential. Specifically, we analyze thermodynamic stability for a single unit cell of 29.64$^\circ$ twisted bilayer graphene with an interlayer interstitial defect. The total number of atoms in the system is 7397. We perform MD simulation of this system at 1000K for $10^5$ time steps. The results are shown in Figure~\ref{fig:geo}. The fast equilibration followed by stable fluctuations of energy, temperature and pressure over this trajectory indicates long-time stability of the MD system. 
{
We further tested the scalability of our MD simulations using a larger commensurate twisted bilayer graphene supercell at $\theta = 45.318^\circ$ containing 14,884 atoms. The results are showin in Figure~\ref{fig:geo_larger} in Appendix~\ref{sec:apd:numerics}.
}

\subsubsection{Dynamical properties}
\label{sec:sub:sub:dyn}
Next we present results on dynamic properties, including diffusion coefficients~\cite{wilke1955correlation} and the velocity auto-correlation function (VACF). For these tests we can no longer compare to reference DFT predictions, hence these tests only confirm the robustness and qualitative correctness of our model for these simulation tasks.

The diffusion coefficient in twisted bilayer graphene is important for understanding its ability to intercalate (e.g., by lithium), and thus access its potential as an energy material candidate~\cite{shen2020correlated}. It can be computed from atomic trajectories obtained via molecular dynamics (MD) simulations using the mean squared displacement (MSD) as a function of time,~\cite{einstein1956investigations} 
\begin{equation}
    D = \lim_{t \to \infty} \frac{\langle | \bm{r}(t) - \bm{r}(0) |^2 \rangle}{6t},
\end{equation}
where $\bm{r}(t)$ is the particle position at time $t$ and angular brackets $\langle \cdot \rangle$ denotes the canonical average, approximated in practice as an average over all particles in the system and multiple samples of time origins along the MD trajectory. 

\begin{figure}
\centering
\includegraphics[height=4.8cm]{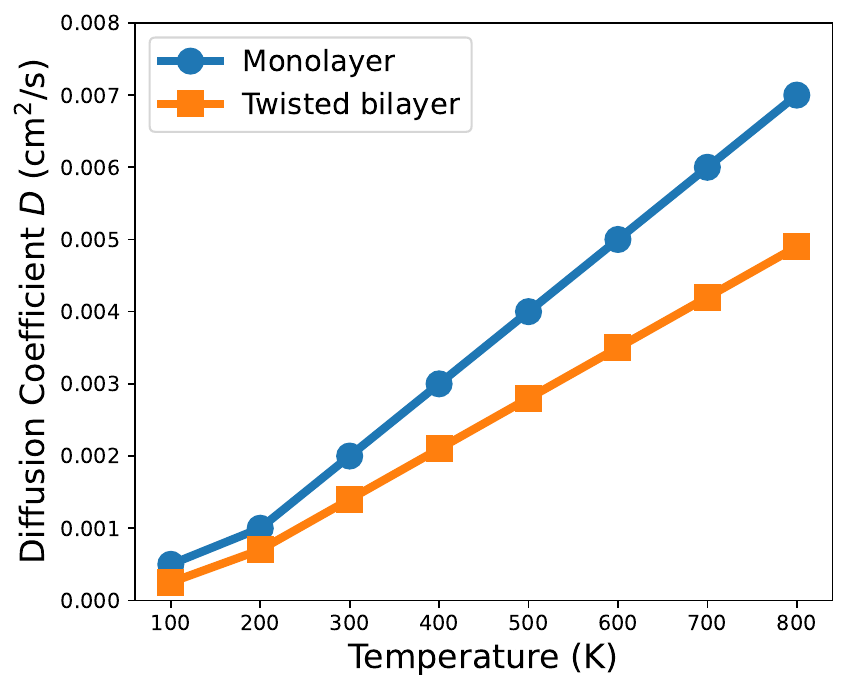}
\includegraphics[height=4.8cm]{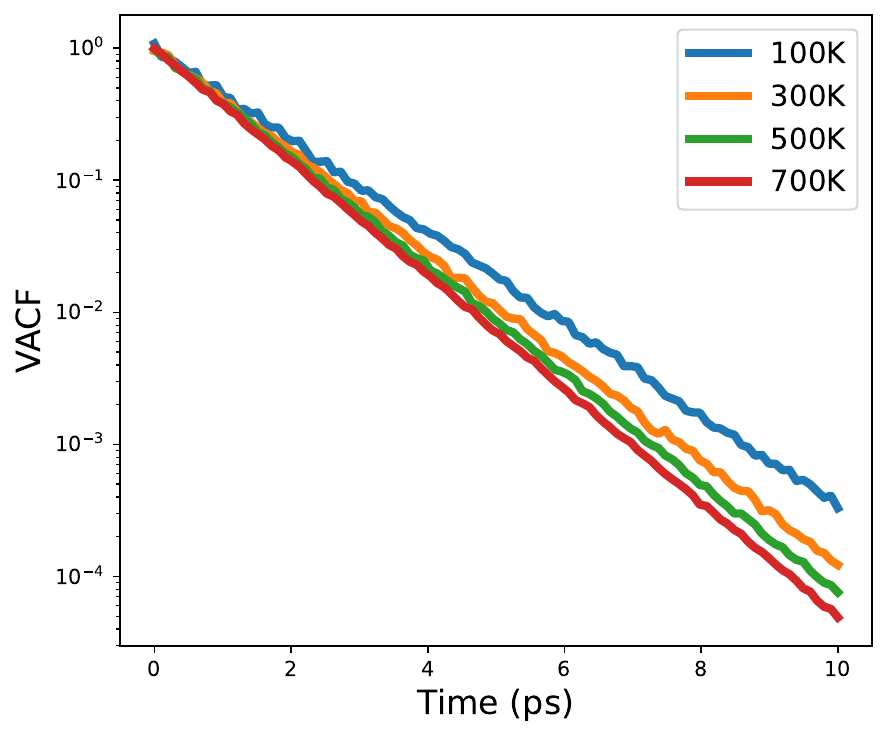}
\includegraphics[height=4.8cm]{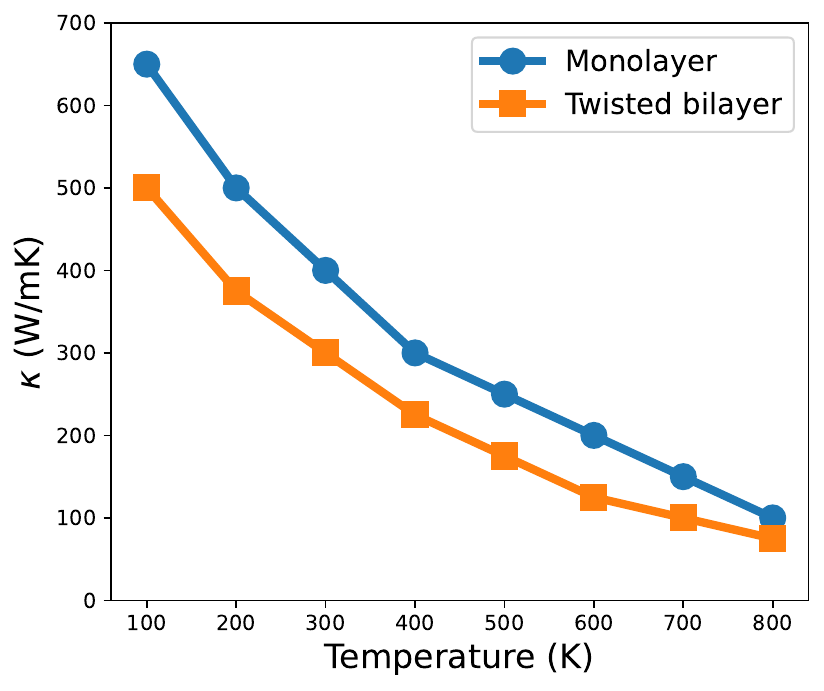}
\caption{{\bf Left:} Diffusion coefficients of monolayer and twisted bilayer graphene for various temperatures. {\bf Center:} The { VACF} of twisted bilayer graphene for various temperatures. 
{\bf Right:} Thermal conductivity at various temperatures with the fitted \ACEstar\ potential for monolayer and twisted bilayer graphene. 
}
\label{fig:diffusion}
\label{fig:thermal}
\end{figure}

We compute the diffusion coefficients using the above formulation and present the results for various temperatures in Figure~\ref{fig:diffusion}. The results align with the expectation that the diffusion coefficient increases with temperature. While~\cite{guerrero2024disorder} provides relevant results that do not account for temperature dependence, to the best of the authors’ knowledge, no reference data or experimental results are currently available for direct comparison.

The { VACF} is another important dynamical property, providing insights into the temporal correlations of particle velocities in twisted bilayer graphene. The VACF $C_v(t)$ can be computed from the velocities obtained from molecular dynamics simulations and is important for understanding thermal transport and the dynamic stability of the material~\cite{cheng2023magic}. The VACF $C_v(t)$ is defined as:
\begin{equation}
C_v(t) = \langle \bm{v}(0) \cdot \bm{v}(t) \rangle,
\end{equation}
where $\bm{v}(t)$ is the velocity vector of a particle at time $t$, and $\bm{v}(0)$ is the initial velocity vector of the particle. The angle brackets denote an ensemble average over all particles and time origins. In practice, the VACF is computed as:
\begin{equation}
C_v(t) = \frac{1}{N} \sum_{i=1}^{N} \frac{1}{T-t} \int_{0}^{T-t} \bm{v}_i(\tau) \cdot \bm{v}_i(\tau + t) \, d\tau,
\end{equation}
where $N$ is the number of particles, and $T$ is the total simulation time. 

In Figure~\ref{fig:diffusion}, we present the { VACF} results for twisted bilayer graphene across various temperatures (100K, 300K, 500K, and 700K), plotted on a semi-logarithmic scale. This semi-log plot clearly shows the exponential decay of the VACF over time, with the decay rate increasing as the temperature rises, indicating a faster loss of velocity correlation at higher temperatures. This trend reflects the enhanced atomic vibrations and thermal motion at elevated temperatures, which lead to quicker decorrelation in particle velocities. 

Although finding direct references for comparison is challenging due to the long-time dynamics required, the shape and decay behavior of the VACF here are qualitatively correct. Such details are difficult to capture with conventional DFT, underscoring the utility and necessity of MLIPs in studying systems like twisted bilayer graphene. 


\label{sec:sub:sub:thermal}

Twisted bilayer graphene shows variations in thermal conductivity due to induced changes in electronic and phonon transport mechanisms~\cite{cao2018unconventional, yankowitz2019tuning}. Compared to single-layer graphene, various twist angles can lead to diverse thermal behavior, highlighting the potential for tailoring graphene's thermal properties through precise structural engineering~\cite{kim2016van}. Thermal conductivity in MD simulations reflects how heat is transferred at the atomic level. The Green-Kubo method~\cite{carbogno2017ab}, based on the fluctuation-dissipation theorem, is the most widely used approach for calculating thermal conductivity from equilibrium MD simulations. It is given by 
\begin{equation}\label{eq:Green-Kubo}
    \kappa = \frac{V}{k_B T^2} \int_0^\infty \langle J(0), J(t) \rangle \, {\rm d}t,
\end{equation}
where $V$ is the system volume, $k_B$ is the Boltzmann constant, $T$ is the temperature, $J(t)$ is the heat flux at time $t$, and $\langle J(0), J(t) \rangle$ is the autocorrelation function of the heat flux. In practice, the integral is computed numerically, with the heat flux derived from the time evolution of atomic velocities and positions.  

In twisted bilayer graphene, these parameters can be significantly altered by the twist angle. { The twist introduces a periodic moiré superlattice, which reduces crystal symmetry and enlarges the effective unit cell. This structural modulation folds the phonon bands and leads to the emergence of additional phonon branches. As a result, more phonon-phonon scattering channels become available, especially for Umklapp and normal processes that limit thermal transport.} Experimental data indicate that the thermal conductivity of TBG is significantly lower than that of monolayer graphene and bilayer graphene. { These enhanced scattering processes reduce the phonon mean free paths and lifetimes, thereby suppressing thermal conductivity.} This phenomenon can be attributed to changes in the Brillouin zone and the emergence of additional phonon branches due to the twisted structure, which enhance phonon Umklapp and normal scattering~\cite{faizal2022thermal, li2014thermal}.

We provide a validation of thermal conductivity computed from the molecular dynamics trajectories in Figure~\ref{fig:thermal}. The results indicate that as the temperature increases, the thermal conductivity significantly decreases due to enhanced phonon-phonon scattering at high temperatures. The findings show good qualitative agreement with the literature~\cite{faizal2022thermal}.

\subsection{Phonon Spectra} 
\label{sec:sub:phonon}
To conclude we assess the ability of the \ACEstar\ model for predicting graphene phonon modes. The general phonon equation of motion is given by~\cite{lu2022} 
\begin{equation}
    D_{\mu\nu\alpha\beta} (\bm{q}) \delta u_{\nu\beta} (\bm q) = \omega^2 \delta  u_{\mu\alpha}(\bm{q}),
\end{equation}
where the subscripts $\mu\nu$ denote the Cartesian coordinates, $\alpha\beta$ refer to atomic indices, $\bm q$ is the phonon momentum, $\omega$ represents the phonon frequency, and $\delta u$ is the phonon displacement field. The dynamical matrix element $D_{\mu\nu\alpha\beta} (\bm q)$ is defined as 
\begin{equation}
D_{\mu\nu\alpha\beta} (\bm q) = \sum_{j} \tilde D_{\mu\nu\alpha\beta} (\bm{R}_j) e^{-i \bm q \cdot \bm R_j }, \qquad \tilde D_{\mu\nu\alpha\beta}(\bm R_j) = \frac{1}{M} \frac{ \partial^2 {E^\mathrm{tot}}(\bf A)}{\partial \bm r_{j\mu\alpha} \partial \bm r_{0\nu\beta}}.
\end{equation}
Here, $ij$ represent supercell indices for a $5\times5$ grid of supercells with the central supercell indexed by $i=0$, $\bf R_j$ are the supercell lattice vectors, ${\bf r}_{i\mu\alpha}$ the atomic positions, $\bf A$ the atomic structure, and $M$ is the mass of the Carbon atom. The dynamical matrix elements are proportional to the derivatives of the forces. 
Phonon mode calculations via { DFT} typically involve either differentiating the forces by displacing atoms or employing Density Functional Perturbation Theory (DFPT). Both methods are computationally demanding. Both classical and machine learning interatomic potential models can be differentiated analytically to second order at relatively low computational cost.

\begin{figure}[ht!]
    \centering
    \includegraphics[width=0.8\linewidth]{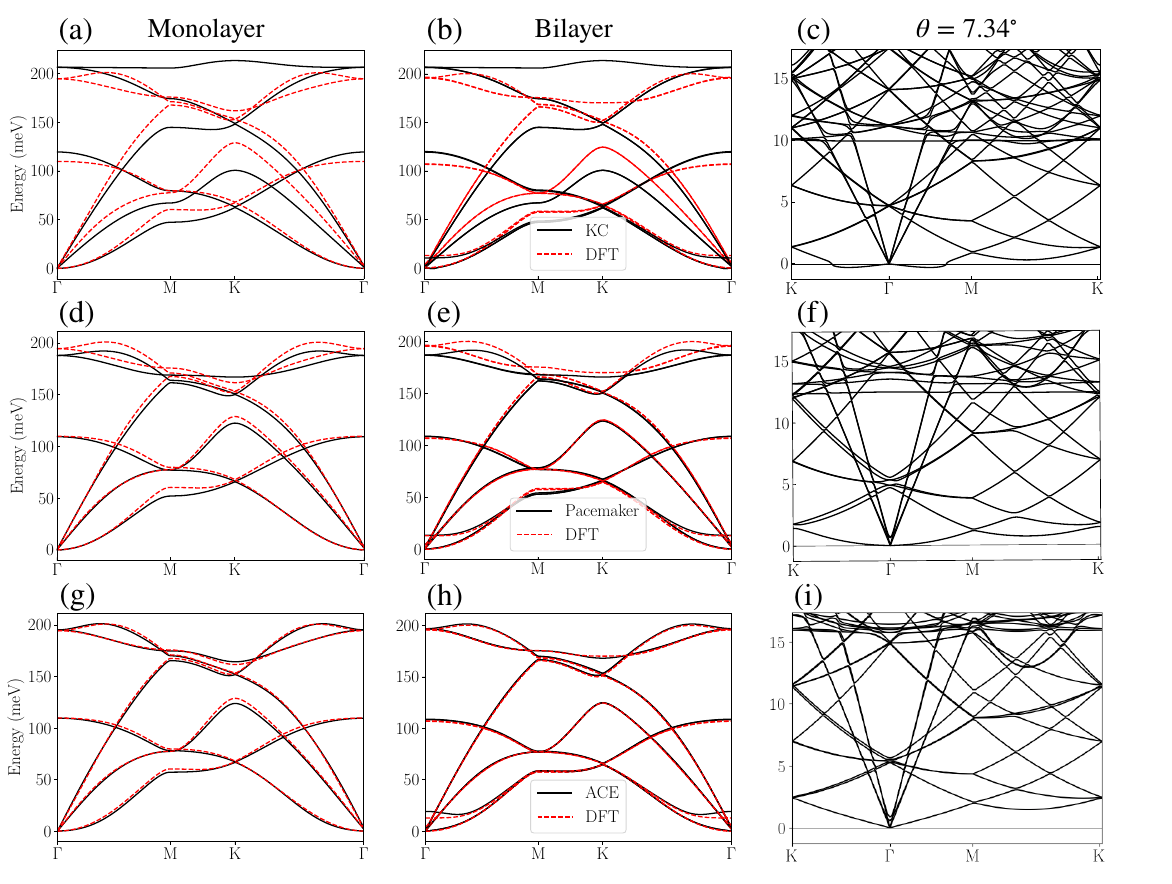}~\hspace{-8mm}~\rotatebox{270}{\hspace{-10.4cm} \small  (a-c) K.-Crespi \hspace{11mm}  (d-f) pacemaker \hspace{9mm} (g-i) \ACEstar } 
    \caption{Phonon band structures obtained from the \ACEstar\ potential (a,b,c), Kolmogorov-Crespi potential (d, e, f) and the general purpose pacemaker C potential (g, h, i).  Black solid lines represent the interatomic potentials generated phonon spectra while red dashed lines represent the DFT-generated results.
    Column (a, d, g) shows the results for monolayer graphene; column (b, e, h) for Bernal/AB-stacked bilayer graphene; column (c, f, i) for twisted bilayer graphene at 7.34$^\circ$ commensurate angle. 
    }
    \label{fig:phonon}
\end{figure}

We evaluate the performance of the fitted \ACEstar\ potential on phonons in Fig.~\ref{fig:phonon}, showing also analogous results for the KC and pacemaker potentials for comparison. Results are shown for monolayer graphene, Bernal/AB-stacked bilayer graphene, and for twisted bilayer graphene at 7.34$^\circ$ commensurate angle. 

For monolayer and for Bernal/AB-stacked bilayer graphene (rigid shift), \ACEstar\ and DFT have an excellent agreement, with the exception of a moderate error in the layer breathing mode near $\Gamma$-point. For these two tests, the KC potential provides good qualitative but not quantitative agreement with DFT. For the optical phonons, both KC and pacemaker potential substantially deviate from the DFT results. 

In the $7.34^\circ$ twisted system, the overall shape of the low-energy phonons, such as folded acoustic phonons, is consistent with DFT predictions (see \cite{lu2022}). 
However, discrepancies appear along the M–K path, and most notably, moir\'e optical phonons near 10~meV are absent. The absence of the optical moir\'e potential is partially due to the overestimate of the layer breathing mode frequency at the $\Gamma$-point in AB bilayer graphene (Fig.~\ref{fig:phonon}(b)). With a twist angle, the moir\'e Brillouin zone becomes much smaller than the monolayer Brillouin zone, which amplifies the discrepancy in the relative frequencies between the folded acoustic modes and the optical modes.

In summary, the custom fitted \ACEstar\ potential significantly outperforms both KC and pacemaker potentials in these tests, but still has limitations in high-energy modes and moir\'e optical phonons in twisted systems. Future work will investigate solutions to these limitations.

\section{Conclusion}
In this study, we have comprehensively explored the use of { MLIPs} to simulate the complex behavior of twisted multilayer graphene, a system characterized by its unique moir\'e patterns. By leveraging the { ACE} architecture, we developed an interatomic potential model that achieves near-DFT (in-distribution) accuracy while significantly reducing computational costs. Our approach introduces a new method to generate training and testing datasets using locally twisted structures capable of exhaustively sampling the large configuration space of twisted multilayered materials. 
%
The model’s performance across a wide range of simulation tasks, including geometry optimizations, molecular dynamics simulations, and phonon calculations, demonstrates the effectiveness of the fitted model both in- and out-of-distribution. Our results demonstrate that the transferrability from non-specialized training datasets to predictions of general twist angles is not guaranteed, but that the locally twisted structures we propose are sufficient to guarantee stable and accurate predictions in a wide range of simulation tasks.

The model is therefore well suited as a starting point for an efficient and robust active-learning approach, adding unseen structures as required for new simulation tasks. It thus provides a theoretical framework for systematically and quantitatively exploring the properties of low-dimensional materials and bridging the gap between microscopic properties and experimental observations such as twist angle disorder~\cite{uri2020} and strain~\cite{kazmierczak2021}, which are frequently encountered but often overlooked in theoretical models.

\section{Acknowledgement}
\label{sec:Ack} 
ML acknowledges support from NSF DMS 2422469.  ML's research was also supported by a visit to the Kavli Institute for Theoretical Physics (NSF PHY-2309135) and by a visit to CCQ at the Simons Flatiron Institute for support while (a portion of) this research was carried out.  DC acknowledges support by DoD NDSEG. VG and SD acknowledge the support from NSF DMS 2422470. CO acknowledges the support of the Natural Sciences and Engineering Research Council of Canada (NSERC) and support through computational resources and services provided by Advanced Research Computing at the University of British Columbia. ZZ is supported by a Stanford Science Fellowship. DM acknowledges support from NSF DMS 2422471.
We finally acknowledge support for a collaboration visit at ICERM.

\appendix

\section{Atomic Cluster Expansion (ACE)}
\label{sec:apd:ACE}
We provide a brief review of the Atomic Cluster Expansion (ACE) model but refer to~\cite{ACE_ralf, DUSSON2022, witt2023otentials} for further details and discussion. Recalling the setup from Section~\ref{sec:sub:ACE}, we consider an atomic configuration \( \mathbf{A} := \{\boldsymbol{r}_1, \ldots, \boldsymbol{r}_N\} \in \mathbb{R}^{3N} \), where $N$ atoms are represented by their position vectors \( \boldsymbol{r}_j \). The relative position vector between atom  $j$ and a reference atom $i$ is \( \boldsymbol{r}_{ij} = \boldsymbol{r}_j - \boldsymbol{r}_i\), and the {\it local} atomic environment around atom $i$ with a cutoff radius $r_{\rm cut}$ is \( \RR_i := \{\boldsymbol{r}_{ij}\}_{j \in \mathcal{N}(i)} \), where $\mathcal{N}(i)$ denotes the set of indices of all atoms within the cutoff radius $r_{\rm cut}$ from atom $i$.

The ACE site energy $\Es(\RR_i; \mathbf{c})$ can be formulated as a body-order expansion for a given correlation order $\nu_{\rm max} \in \mathbb{N}$,
\begin{equation}
\Es(\RR_i; \mathbf{c}) = 
\sum_{\nu=0}^{\nu_{\rm max}} \frac{1}{\nu!} \sum_{j_1, \dots, j_{\nu} \in \mathcal{N}(i)} V_{\nu}(\boldsymbol{r}_{i j_1}, \ldots, \boldsymbol{r}_{i j_{\nu}}; \mathbf{c}),
\end{equation}
where the $(\nu+1)$-body potential $V_{\nu} : \mathbb{R}^{3\nu} \rightarrow \mathbb{R}$ is approximable by using a tensor product basis~\cite{DUSSON2022},
\begin{equation*}
\phi_{{\bm n} {\bm \ell} {\bm m}}(\boldsymbol{r}_{ij_1}, \ldots, \boldsymbol{r}_{ij_{\nu}}) := \prod_{\alpha=1}^{\nu} \phi_{{{\bm n}_{\alpha} {\bm \ell}_{\alpha} {\bm m}_{\alpha}}}(\boldsymbol{r}_{ij_{\alpha}}),
\end{equation*}
and the one-body basis is given by $\phi_{n \ell m}(\boldsymbol{r}) := P_n(y)Y_{\ell}^{m}(\hat{\boldsymbol{r}})$, with $\boldsymbol{r} \in \mathbb{R}^d$, $r = |\boldsymbol{r}|$, and $\hat{\boldsymbol{r}} = \boldsymbol{r}/r$. The functions $Y_{\ell}^{m}$ denote the complex spherical harmonics for $\ell = 0,1,\ldots$, and $m=-\ell,\ldots,\ell$, and $P_n$ are radial basis functions for $n=0,1,\ldots$. 
%
This parametrization is already invariant under permutations (or, relabeling) of the input local atomic environment $\RR_i$. One then symmetrizes the tensor product basis with respect to the group $O(3)$, employing the representation of that group in the spherical harmonics basis. This results in a linear parametrization of the terms of correlation order $\nu$
\begin{align*}
\varepsilon(\RR_i; \mathbf{c})
    &= \sum_{{{\bm n} {\bm \ell} q}} c_{{{\bm n} {\bm \ell} q}} {\bf B}_{{\bm n} {\bm \ell} q} (\RR_i), \\ 
    \text{where} 
    \quad     
    {\bf B}_{{\bm n} {\bm \ell} q} (\RR_{i})
    &= \sum_{{\bm m } \in \mathcal{M}_{ {\bm \ell}}} \mathcal{C}^{{\bm n} {\bm \ell} q}_{{\bm m }} {\bm A}_{{\bm n} {\bm \ell}{\bm m} }(\RR_i), \\ 
    \text{and} 
    \quad 
    {\bm A}_{{\bm n} {\bm \ell}{\bm m} }(\RR_i) &= \prod_{\alpha=1}^{\nu} \sum_{j\in \mathcal{N}(i)} \phi_{{\bm n}_{\alpha}{\bm \ell}_{\alpha}{\bm m}_{\alpha}}(\boldsymbol{r}_{ij}).
\end{align*}


In the above, the $q$-index enumerates the number of all possible invariant couplings through the generalized Clebsch--Gordan coefficients $\mathcal{C}^{{\bm n} {\bm \ell} q}_{{\bm m }}$, $\mathcal{M}_{{\bm \ell}}:=\{{\bm m}\in\mathbb{Z}^{\nu}|-\ell_{\alpha} \leq {\bm m}_{\alpha} \leq \ell_{\alpha}\}.$ 

This parametrization was originally devised in \cite{ACE_ralf}. 
The implementation we employ in our work is described in \cite{witt2023otentials}. The latter reference also described the details of the choice of radial basis $P_n$.

To complete the description of our parametrization, we select a finite subset of the basis, 
\begin{align*}
\mathcal{B} := \Big\{ {\bf B}_{{\bm n} {\bm \ell} q} \,\Big|\, & ({\bm n}, {\bm \ell}) \in \mathbb{N}^{2\nu} \text{ ordered}, ~ \sum_{\alpha} {\bm \ell}_{\alpha} \text{ even}, ~  
    \sum_{\alpha} {\bm m}_\alpha = 0, \\ 
    & q = 1, \ldots, n_{{\bm n}{\bm \ell}}, ~ 
    \sum_\alpha {\bm \ell}_\alpha + {\bm n}_\alpha \leq D_{\rm tot}, ~
    \nu \leq \nu_{\rm max} 
    \Big\},
\end{align*}
where the two approximation parameters are the correlation order $\nu_{\rm max}$ and the total degree $D_{\rm tot}$. To facilitate the use of permutation invariance, we define $({\bm n}, {\bm \ell}) \in \mathbb{N}^{2\nu}$ as ordered if the vector associated with this tuple is lexicographically ordered~\cite[Definition 1]{DUSSON2022}. Additionally, we impose the condition that $\sum_{\alpha} {\bm \ell}_{\alpha}$ must be even to ensure point reflection symmetry. The constraint $\sum_{\alpha} {\bm m}_\alpha = 0$ arises from the requirement that rotationally invariant basis functions are constructed using specific linear combinations of spherical harmonic products, which couple to a total angular momentum of zero. This property is derived purely within the framework of linear algebra. For a more detailed discussion, we refer the reader to~\cite[Section 3.4]{DUSSON2022}. With this selection of the basis we can write our ACE parametrization more conveniently as 
\begin{equation}
\Es(\RR_i; \mathbf{c}) = \sum_{{\bf B} \in \mathcal{B}} c_{\bf B} {\bf B}(\RR_i), 
\end{equation}
where ${\bf c} = \{c_{\bf B}\}_{{\bf B} \in \mathcal{B}}$ are the model parameters. 
%
Due to the completeness of this representation, as the approximation parameters (correlation order $\nu_{\rm max}$, cut-off radius $r_{\rm cut}$, and total degree $D_{\rm tot}$) approach infinity, the model can {\it in principle} represent any arbitrary potential~\cite{DUSSON2022}.

Due to the rapid developments in MLIPs architectures and software, performance benchmarks would quickly become outdated. We therefore chose to not perform such benchmarks as part of our study. However, we briefly mention that the implementation we used at the time of writing has an inference cost of approximately 100$\mu$s per atom for mono-layer graphene and approximately 200$\mu$s per atom for bilayer graphene in single-threaded execution on a AMD EPYC-Rome (2000 MHz) CPU. We expect that designing a multi-layer graphene MLIP with primarily performance optimizations in mind can achieve comparable accuracy with significantly improved performance characteristics.

\section{Bayesian Linear Regression}
\label{sec:apd:blr}

In this section, we discuss the Bayesian linear regression that originally comes from the minimization of the loss (cf.~\eqref{eq:quad-loss}). Since all observations we consider here are linear, the minimization of $\mathcal{L}(\mathbf{c})$ can be written in the form of the ridge regression
\begin{equation}
\label{eq:ridge}
\underset{\mathbf{c}}{\mathrm{argmin}} \, \| \mathbf{W} (\mathbf{O} - \mathbf{\Psi}\mathbf{c}) \|^2 + \lambda \|\mathbf{\Gamma} \mathbf{c}\|^2,
\end{equation}
where $\mathbf{O}$ is a vector containing the observation values $E^{\rm ref}$, $F^{\rm ref}$, and $V^{\rm ref}$.$\mathbf{\Psi}$ is the design matrix containing the ACE basis values corresponding to those observations, and $\mathbf{W}$ is a diagonal matrix containing the weights $w_{E,\mathbf{R}}$ and $w_{F,\mathbf{R}}$. $\mathbf{\Gamma}$ specifies the form of the regularizer and $\lambda$ is a scaling parameter determining the relative weight of the regularization. This formulation of the least squares problem is often also called regularized least squares, and the $\lambda \|\mathbf{\Gamma} \mathbf{c}\|^2$ term is often called generalized Tikhonov regularization. The default for $\mathbf{\Gamma}$ is zero or the identity, depending on the choice of solver. Our recommendation is to use the smoothness prior introduced in~\cite{witt2023otentials} instead for most solvers.

Uncertainty estimates of model predictions are highly sought after tools to judge the accuracy of a prediction during simulation with a fitted model, but can also be employed to great effect during the model development workflow, e.g., in an active learning context. Such uncertainty estimates can be derived in a principled way by recasting the minimization problem~\eqref{eq:quad-loss} in a Bayesian framework where inference is based on the Bayesian posterior distribution
\begin{equation}\label{eq:post}
\text{post}(\mathbf{c}) := p(\mathbf{c}|\mathbf{\Psi}, \mathbf{O}) \propto p(\mathbf{ \mathbf{O}} | \Psi,\,\mathbf{c}) p(\mathbf{c}).
\end{equation}
Here, $p(\mathbf{ \mathbf{O}|\Psi},\mathbf{c})$ denotes the likelihood of the observed data, and $p(\mathbf{c})$ the prior distribution on the model parameters. The Bayesian analogue of \eqref{eq:quad-loss} is a Bayesian Linear Regression model with Gaussian observational noise and prior
\begin{align}
p(\mathbf{ \mathbf{O} | \Psi}, \mathbf{c}) &\propto \exp \left( -\frac{1}{2} (\mathbf{O} - \mathbf{\Psi}\mathbf{c})^\mathrm{T} (\beta \mathbf{W}^2)(\mathbf{O} - \mathbf{\Psi}\mathbf{c}) \right), \\ \nonumber
p(\mathbf{c}) &\propto \exp \left( -\frac{1}{2} \mathbf{c}^\mathrm{T} \mathbf{\Sigma}_0^{-1} \mathbf{c} \right),
\end{align}
where the covariance $\beta^{-1} \mathbf{W}^{-2}$ of the observation noise depends on the regression weight matrix $\mathbf{W}$ defined in~\eqref{eq:ridge} and a hyperparameter $\beta > 0$. $\Sigma_0$ is the covariance matrix of prior $p(\mathbf{c})$. This choice of prior and noise model yields a Gaussian posterior distribution, $p(\mathbf{c} | \mathbf{\Psi}, \mathbf{O}) = \mathcal{N}(\mathbf{c}; \bar{\mathbf{c}}, \mathbf{\Sigma})$, with mean and covariance given, respectively, by \( \bar{\mathbf{c}} = \beta \mathbf{\Sigma} \mathbf{\Psi}^\mathrm{T} \mathbf{W}^2 \mathbf{O} \) and \( \mathbf{\Sigma} = \left( \beta \mathbf{\Psi}^\mathrm{T} \mathbf{W}^2 \mathbf{\Psi} + \mathbf{\Sigma}_0^{-1} \right)^{-1} \). The above Bayesian model can be connected to the ridge regression formulation of~\eqref{eq:ridge} by noticing that maximising the posterior density~\eqref{eq:post} is equivalent to minimizing the regularized loss in~\eqref{eq:ridge} when $\Sigma_0^{-1}=\zeta\mathbf{\Gamma}^2$ for some $\zeta>0$ and $\lambda=\zeta/\beta$. 

The reliability of uncertainty estimates critically depends on the values of the model hyper-parameters, the noise and prior covariance matrices \( \beta^{-1}\mathbf{W}^{-2} \) and \( \mathbf{\Sigma}_0 \). In ACE, it is sometimes difficult to make informed guesses of explicit values of these hyper-parameters that lead to good fits. We therefore commonly employ empirical Bayes approaches that infer appropriate values of these parameters directly from the training data by virtue of maximizing the model evidence.
\begin{align}
p(\mathbf{ \mathbf{O}|\Psi},\mathbf{\Sigma}_0, \beta) &= \int p( \mathbf{O} | \mathbf{\Psi}, \mathbf{c}, \beta) p(\mathbf{c} | \mathbf{\Sigma}_0) d\mathbf{c} \nonumber  \\ 
&= \sqrt{\frac{\beta (2\pi)^{-N_{\text{obs}}} \left|\mathbf{\Sigma}\right|}{\left|\mathbf{\Sigma}_0\right| \left|\mathbf{W}^{-2}\right|}} \exp\left( -\frac{1}{2} (\mathbf{O} - \mathbf{\Psi}\bar{\mathbf{c}})^\mathrm{T} (\beta \mathbf{W}^2)(\mathbf{O} - \mathbf{\Psi}\bar{\mathbf{c}}) - \frac{1}{2} \bar{\mathbf{c}}^\mathrm{T} \mathbf{\Sigma}_0^{-1} \bar{\mathbf{c}} \right)
\end{align}
as a function of \( \mathbf{\Sigma}_0, \beta \). Intuitively, maximizing the model evidence results in a model where the regularizing effect of the covariance matrix \( \mathbf{\Sigma}_0 \) and the degree of penalization of model misfit—modeled by the noise covariance matrix \( \beta^{-1}\mathbf{W}^{-2} \)—are balanced against the degree to which the regression coefficients are determined by the data.

The Bayesian ridge solver provides a posterior parameter distribution $p(\mathbf{c})$. It remains important to estimate the uncertainty of predictions. Evaluating such uncertainties from the exact posterior distribution is computationally expensive; so one can draw $K$ samples $\{\mathbf{c}_k\}_{k=1}^K$ from $\text{post}(\mathbf{c})$ resulting in a committee of ACE models which can be used to obtain computationally efficient uncertainty estimates for predictions. For example, the standard deviation $\sigma$ of a total energy prediction can be approximated by a committee via
\begin{equation}
\sigma^{\rm E} = \sqrt{\frac{1}{\beta w_{E,R}^2} + \frac{1}{K} \sum_{k=1}^K (E^k - \overline{E})^2},
\end{equation}
where $\overline{E}:=\sum_{i=1}^{N}\sum_{\mathbf{B} \in \mathcal{B}} \bar{\mathbf{c}} \cdot \mathbf{B}(\RR_i)$ with $\bar{\mathbf{c}}$ 
 being the posterior mean of the distribution and $E^k$ denoting the total ACE energy evaluated using $\mathbf{c}^k$.
Similarly, the corresponding relative force uncertainty, as defined in \cite{hyperactive2022}, is given by:
\begin{equation*}
\sigma^{\rm F}_i := \frac{1}{K} \sum_{k=1}^{K} \frac{\| F_i^k - \overline{F}_i \|}{\|\overline{{F}}_i\| + \varepsilon},
\end{equation*}
where $\overline{{F}}_i$ is the mean force prediction, and $\varepsilon$ is a regularization constant to prevent divergence, typically set to around 0.2 eV/Å. 

\section{Training Details}
\label{sec:apd:supp}

\begin{figure}
\centering
\includegraphics[height=5.5cm]{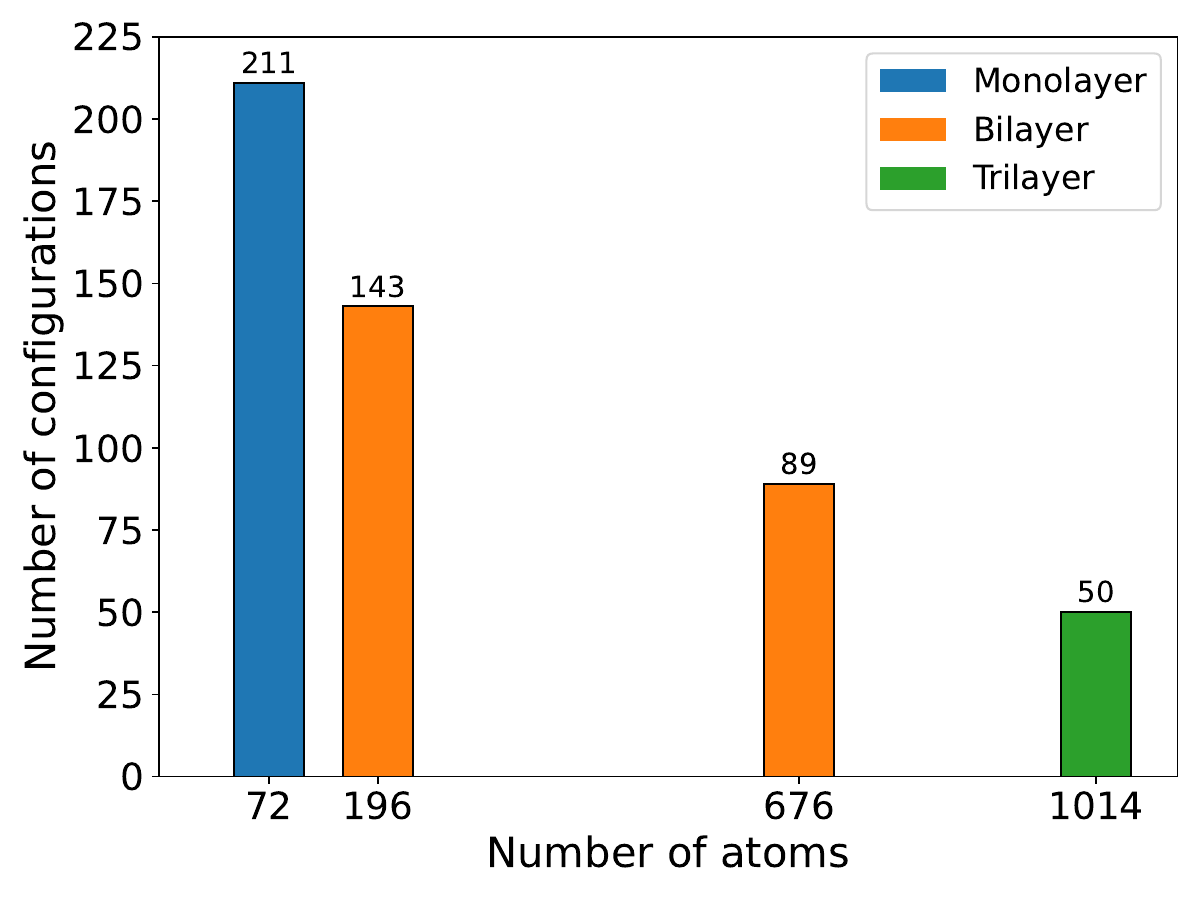}~~
\includegraphics[height=5.5cm]{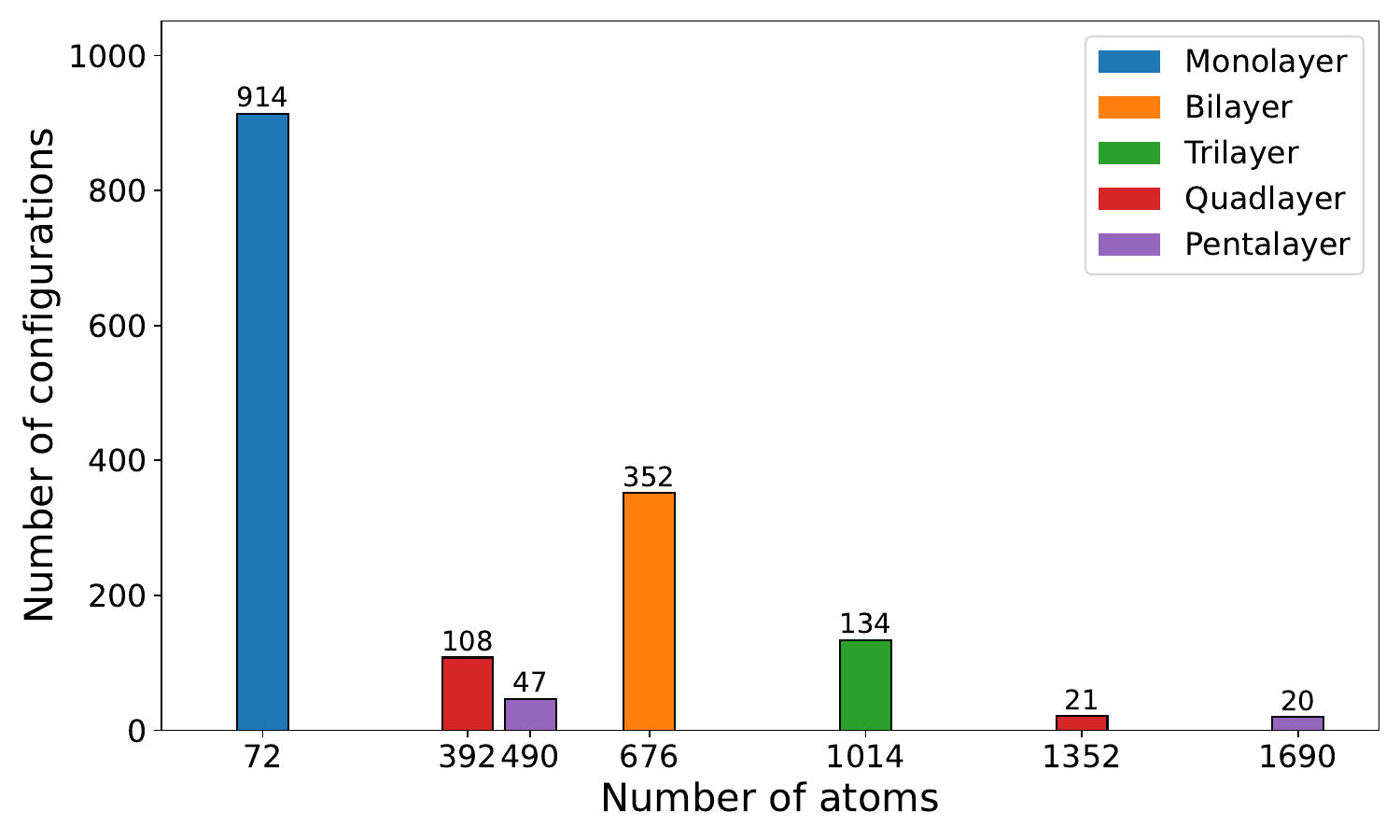}
\caption{
Distribution of configuration counts with respect to the number of atoms in the training and test sets. 
Left: training set used for ACE model fitting; 
Right: test set corresponding to the results reported in Table~\ref{tab:fittingaccuracy}.
}
\label{fig:datasets}
\end{figure}

The DFT computations of the graphene layers in this work were performed using the \texttt{DFT-FE} software~\cite{das2022dft,MOTAMARRI2020106853,Motamarri2018,MOTAMARRI2013308}, a massively parallel open-source software for real-space Kohn-Sham DFT calculations using a systematically convergent adaptive spectral finite-element discretization. DFT-FE employs the Chebyshev filtering acceleration technique~\cite{ChFSISAAD2006,MOTAMARRI2013308} for solving the Kohn-Sham eigenvalue problem, which in conjunction with mixed precision algorithms~\cite{GB2019} provides a computational complexity for DFT-FE that scales quadratically with the number of electrons up to system sizes of 30,000 electrons~\cite{MOTAMARRI2020106853}. The forces and stress tensor are computed using the configurational forces approach ~\cite{Motamarri2018}. Further, the GPU acceleration of the code enables fast and accurate large-scale DFT calculations using DFT-FE~\cite{das2022dft,GB2019,GB2023}. The DFT training data was generated on system sizes up to 1014 atoms, the largest systems corresponding to trilayer graphene. The largest system sizes for which the DFT data was obtained for testing included up to 1352 atoms for quadrilayer graphene and 1690 atoms for pentalayer graphene. { We employ the PBE exchange-correlation functional and the optimized norm-conserving Vanderbilt pseudopotentials (ONCV)~\cite{oncv2013} from the SPMS library~\cite{SHOJAEI2023108594}. All numerical parameters in DFT-FE were chosen such that the ground-state energies and interatomic forces are converged to an accuracy of under 3 meV/atom and 8 meV/\AA, respectively. Additionally, we used a Fermi-Dirac smearing with a temperature of 500 K for all simulations. We note that the DFT-D3~\cite{DFTD3} long-range dispersion corrections are added separately to the ACE model trained only with PBE exchange-correlation functional-based data.} Figure~\ref{fig:datasets} shows the distribution of configuration counts as a function of the number of atoms in the training and test sets.

{
Although this study employs DFT-PBE data, our method is compatible with higher-level reference data (e.g., HSE06, GW), and can be extended accordingly when such data becomes available.
}

\section{Dataset Disregistry Generation}
\label{sec:apd:disr}

\begin{figure}[h]
\centering
\includegraphics[height=6.5cm]{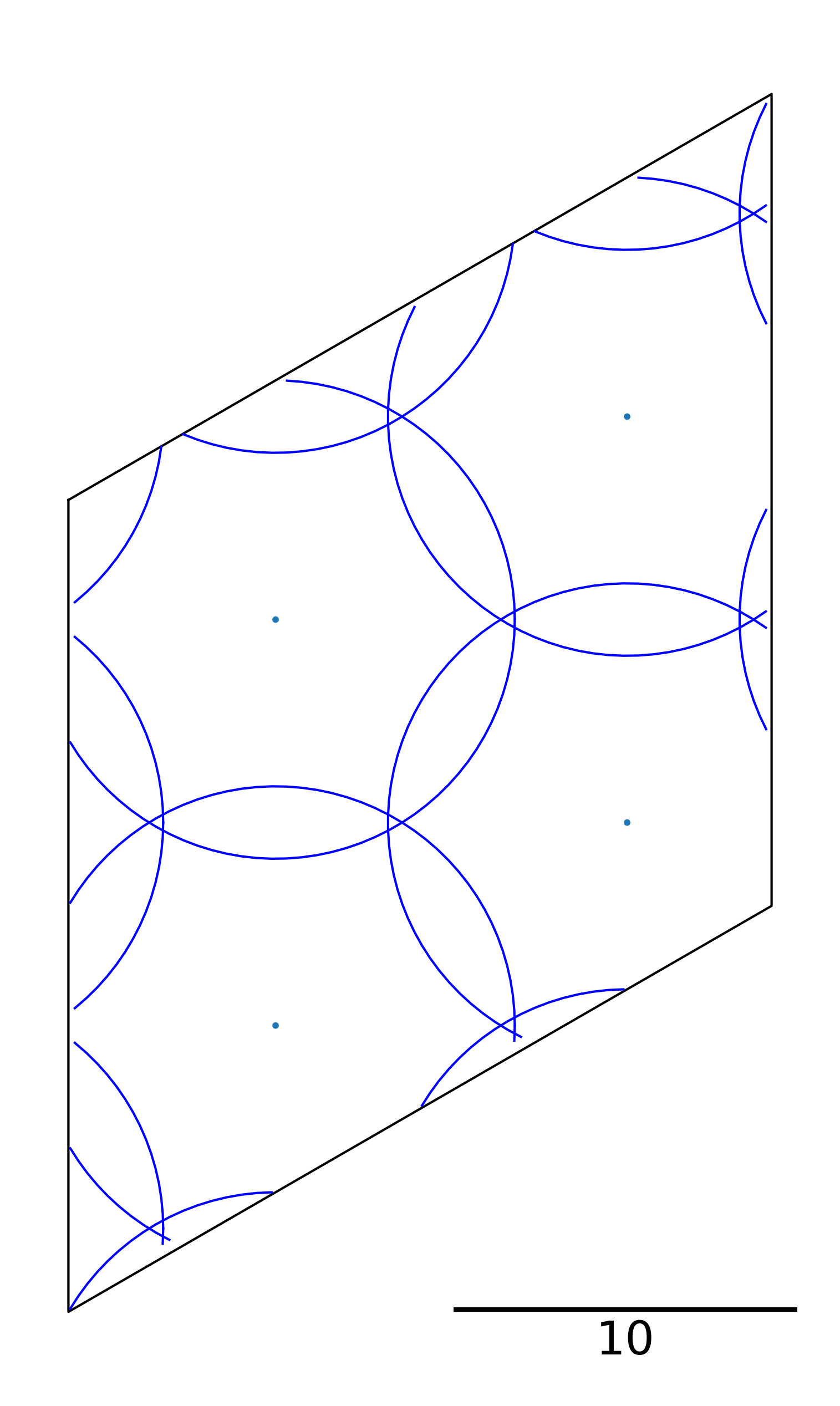} \quad
\includegraphics[height=6.5cm]{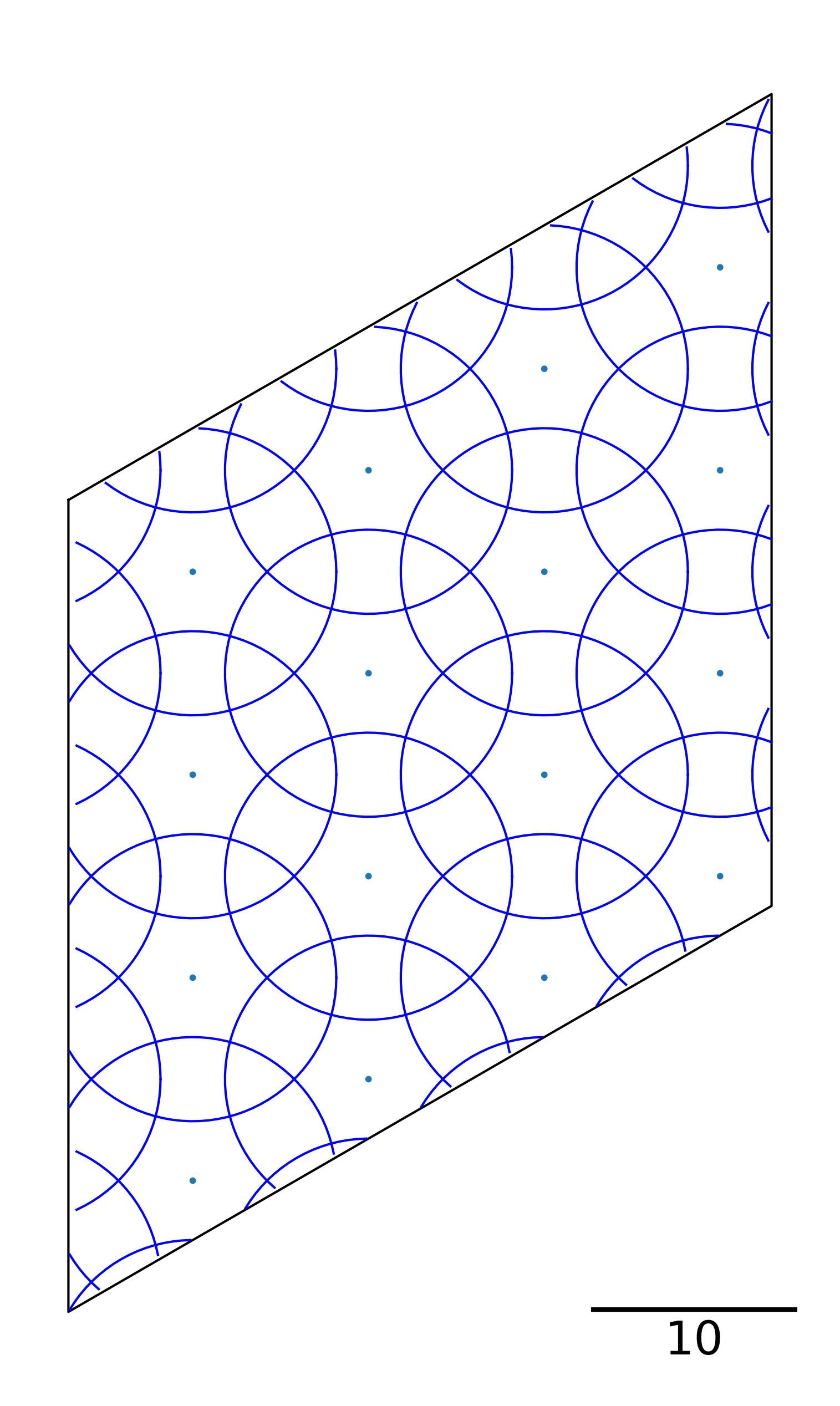} \quad
\includegraphics[height=6.5cm]{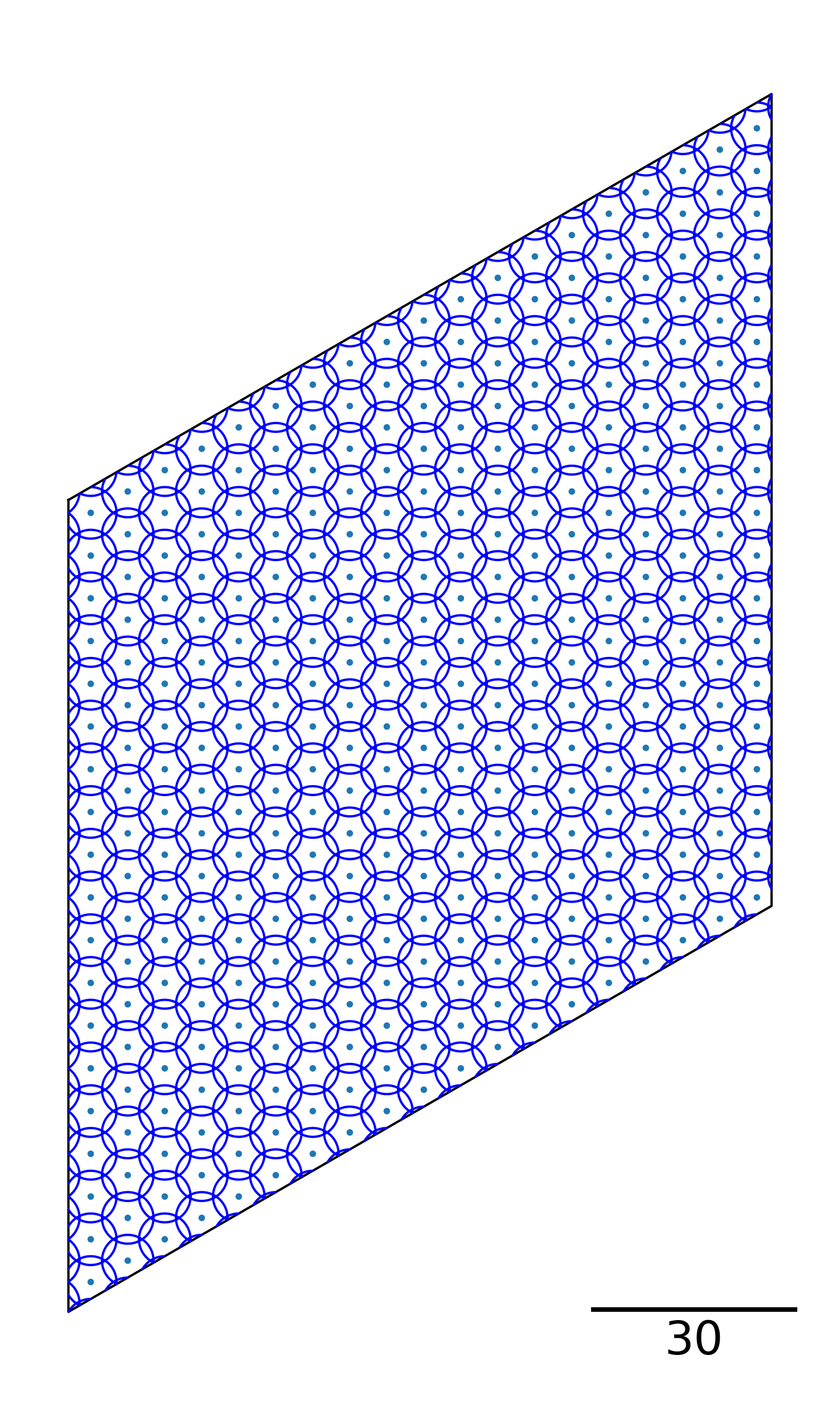}
\caption{Moir\'e unit cell with covering $r_1$ radius circles centered on disregistry points chosen for various twist angles: $6.0^\circ$ (left), $3.6^\circ$ (middle), $1.2^\circ$ (right).}
\label{fig:disregistry_spacing}
\end{figure}

The choice of the disregistry in the local twisting method allows us to sample any point in the moir\'e unit cell.  The local twisting method produces a structure with a central disk of radius $r_1$ that matches the configuration of an $r_1$ radius disk in the moir\'e cell.  The location of this corresponding disk in the moir\'e cell is determined by the disregistry used in the local twisting method.  So to ensure the entire moir\'e cell is present in the locally twisted configurations we just have to pick disregistries such that disks of radius $r_1$ centered on each disregistry cover the moir\'e cell, i.e., our set of disregistries must be an $r_1$-covering of the moir\'e cell.  Since the moir\'e cell size grows as the angle decreases, more disregistries are needed to cover smaller local twist angles than larger ones.  The procedure we used was as follows: pick a starting point $p_1$ on a diagonal of the moir\'e cell that is $r_1$ distance away from a corner.  Find the minimum spacing needed so that an evenly spaced grid including $p_1$ aligned with the boundaries of the unit cell covers the cell.  Some example disregistry grids generated with this method for various angles are shown in Figure \ref{fig:disregistry_spacing}.

The data and code will be made available upon publication via a GitHub repository.

{

\section{Numerical Supplement}
\label{sec:apd:numerics}

We tested the stability of MD simulations at 1000K for $4.41^\circ$ twisted bilayer graphene using two ACE models: one trained on globally twisted data (ACE~(gt)) and the other on locally twisted data (ACE~(lt)). As shown in Figure~\ref{fig:md}, the ACE~(gt) model becomes unstable after 2500 steps, with temperature and energy diverging. In contrast, ACE~(lt) remains stable for 20,000 steps, highlighting the importance of locally twisted configurations for robust simulations.

\begin{figure}
\centering
\includegraphics[height=12cm]{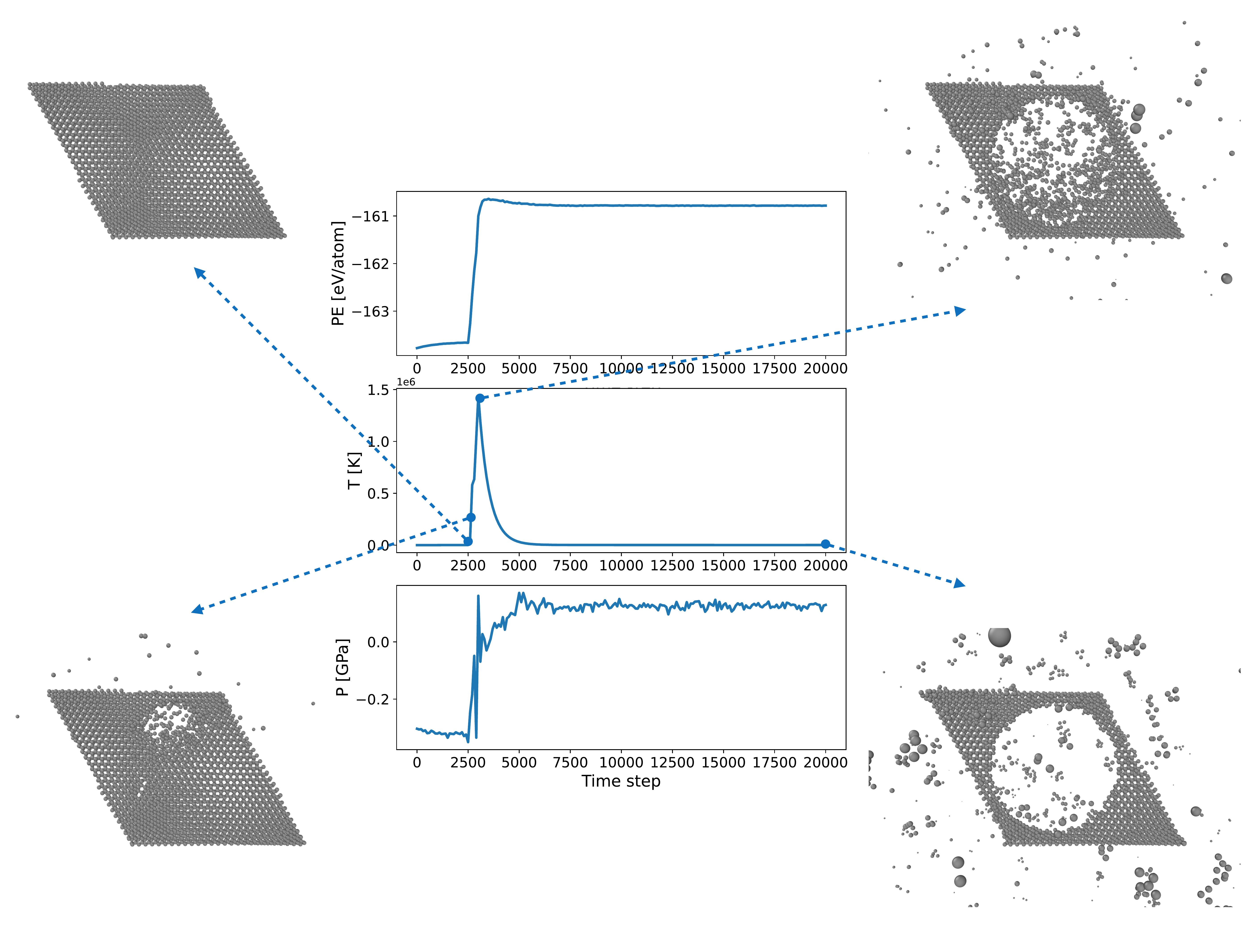}
\includegraphics[height=9cm]{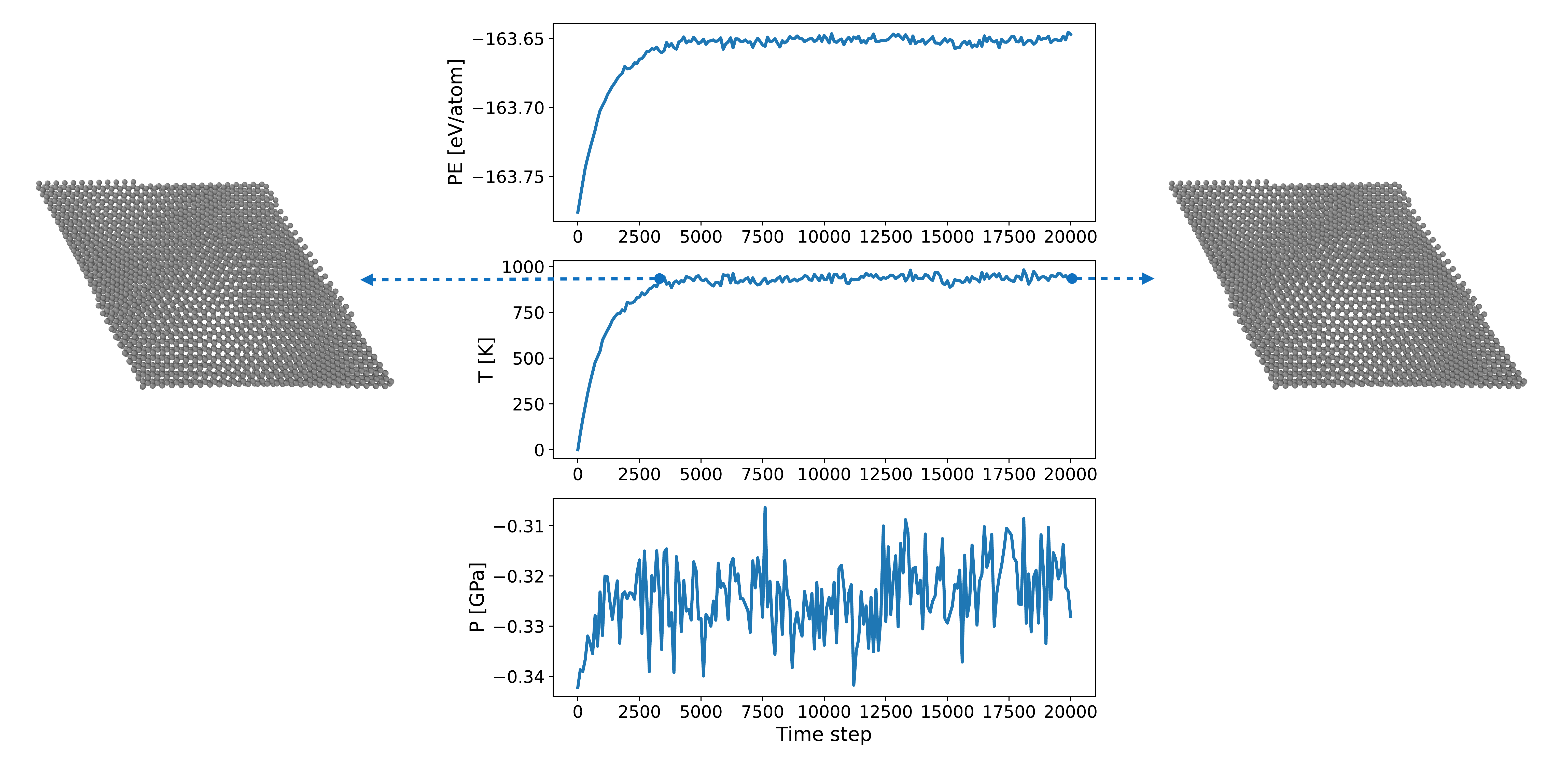}
\caption{Trajectories and thermodynamic properties from molecular dynamics simulations of $4.41^\circ$ twisted bilayer graphene at 1000K using the ACE (gt) potential (Top) and the ACE (lt) potential (Bottom). The ACE (gt) potential trained only on globally twisted results in unstable behaviour, while dynamics under potential ACE (lt) remains stable for the entire trajectory.}
\label{fig:md}
\end{figure}

To assess the scalability of our MLIP in large-scale simulations, we performed a molecular dynamics run on a commensurate bilayer graphene system with a twist angle of $45.318^\circ$, containing 14,884 atoms and an interlayer interstitial defect. The simulation was carried out at 1000~K using a Langevin thermostat with DFT-D3 dispersion corrections and a time step of 1~fs for 5000 steps. The system remained stable throughout the trajectory, demonstrating the robustness of the \ACEstar model. 

\begin{figure}
\centering
\includegraphics[height=10cm]{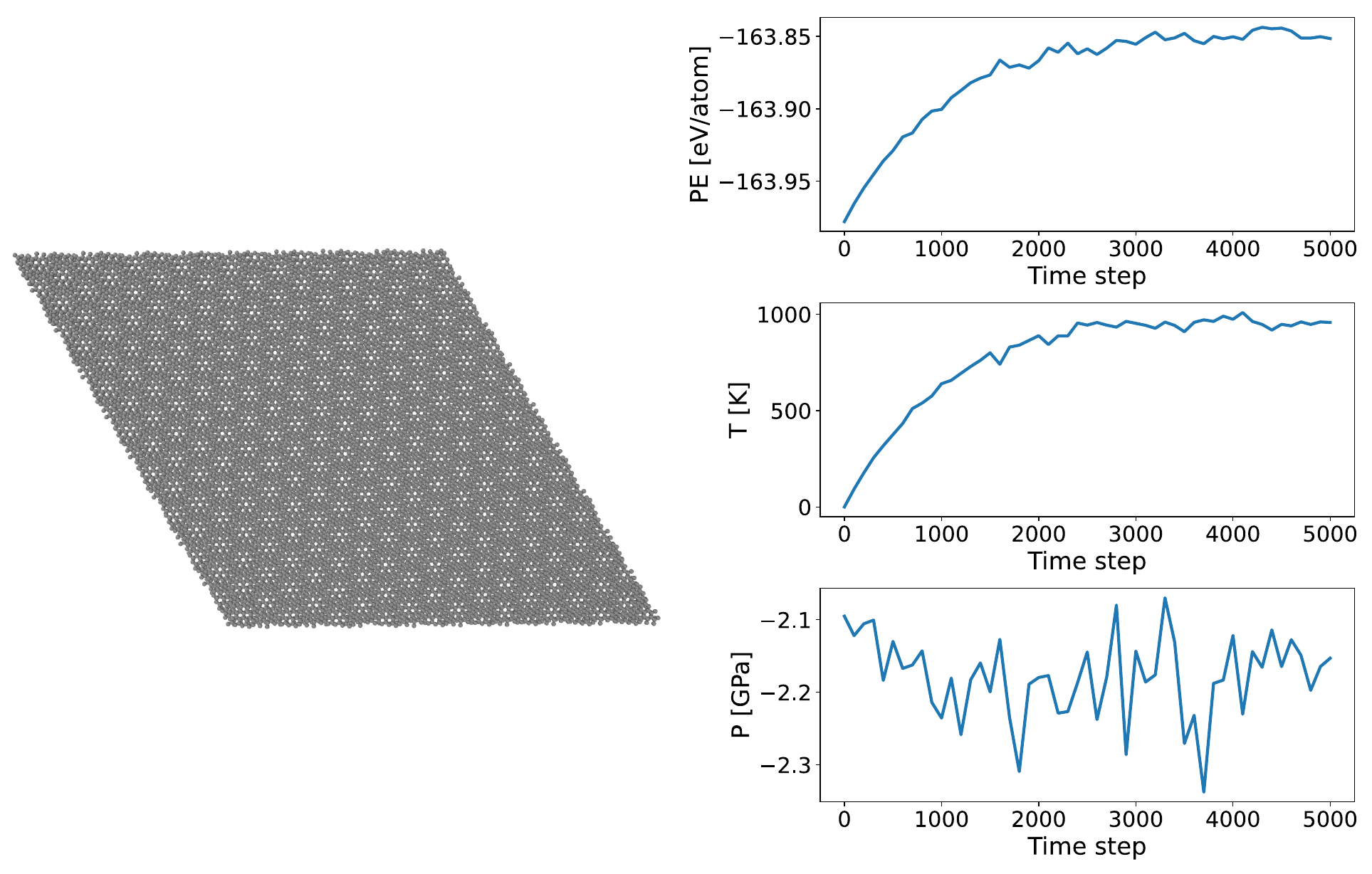}
\caption{{\bf Left:} Atomic configuration of a commensurate bilayer graphene structure with a twist angle of $45.318^\circ$, containing 14,884 atoms. {\bf Right:} Evolution of total energy during a 1000~K molecular dynamics simulation over 5000 steps.}
\label{fig:geo_larger}
\end{figure}

}


\bibliographystyle{unsrt}
\bibliography{mlips}

\end{document}